\documentclass[aps,twocolumn,showpacs,superscriptaddress,groupedaddress]{revtex4}  
\usepackage{graphicx}  
\usepackage{dcolumn}   
\usepackage{bm}        
\usepackage{amssymb}   
\usepackage{amsmath}
\usepackage{mathtools}

\usepackage{color}

\newcommand{\be}{\begin{equation}}
\newcommand{\ee}{\end{equation}}
\newcommand{\beq}{\begin{eqnarray}}
\newcommand{\eeq}{\end{eqnarray}}

\def\barnue{\mathrel{{\bar \nu}_e}}

\def\msun {$M_{\odot}$}

\newcommand{\sn}{supernova}

\newcommand{\n}{neutrino}
\newcommand{\ns}{neutrinos}
\newcommand{\gw}{GW}

\newcommand{\ck}{Cherenkov}
\newcommand{\hk}{Hyper-K}
\newcommand{\ic}{IceCube}
\newcommand{\si}{SASI}
\newcommand{\nsi}{no-SASI}
\newcommand{\kur}{KKHT}
\newcommand{\cmr}{Cramer-Rao}
\newcommand{\fm}{FIM}

\definecolor{mypink1}{rgb}{0.858, 0.188, 0.578}
\definecolor{mygreen}{rgb}{0, 0.4, 0.07}
\definecolor{mypurple}{rgb}{0.5, 0, 0.85}

\begin{document}

\title{Detectability of SASI activity in supernova neutrino signals }

\author{Zidu Lin }

\author{Cecilia Lunardini}

\affiliation{Department of Physics, Arizona State University, \\ 450 E. Tyler Mall, Tempe, AZ 85287-1504 USA}

\author{Michele Zanolin}
\affiliation{Embry-Riddle Aeronautical University, Prescott, Arizona 86301, USA}

\author{Kei Kotake}
\affiliation{Department   of   Applied   Physics \& Research Institute of Stellar Explosive Phenomena, Fukuoka University,    8-19-1,Nanakuma, Fukuoka, 814-0180, Japan}

\author{Colter Richardson}
\affiliation{Embry-Riddle Aeronautical University, Prescott, Arizona 86301, USA}

\date{\today}

\begin{abstract}
We introduce a novel methodology for establishing the presence of Standing Accretion Shock Instabilities (SASI) in the dynamics of a core collapse supernova from the observed neutrino event rate at water- or ice-based neutrino detectors. The methodology uses a likelihood ratio in the frequency domain as a test-statistics; it is also employed to assess the potential to estimate the frequency and the amplitude of the SASI modulations of the neutrino signal. The parameter estimation errors are consistent with the minimum possible errors as evaluated from the inverse of the Fisher information matrix, and close to the theoretical minimum for the SASI amplitude. Using results from a core-collapse simulation of a 15 solar-mass star by Kuroda {\it et al.} (2017) as a test bed for the method, we find that SASI can be identified with high confidence for a distance to the supernova of up to $\sim 6$ kpc for IceCube and and up to $\sim 3$ kpc for a 0.4 Mt mass water Cherenkov detector. This methodology will aid the investigation of a future galactic supernova.

\end{abstract}

\maketitle

\section{Introduction}

The numerical study of the dynamics of core-collapse supernovae allowed in the recent decades to identify specific hydrodynamics mechanisms which control the evolution of the shock wave. Among these dynamics, one that is expected to produce signatures both in the neutrino luminosity and the gravitational wave emission is the Standing Accretion Shock Instability (SASI) \cite{Blondin:2002sm,Foglizzo:2015dma}. 
SASI is a hydrodynamical mode with a typical
frequency,  phase and possibly varying amplitude that develops when a deformed stalled shock front precesses around the newly formed proto-neutron star (PNS). Such precession in turn induces an asymmetric accretion onto the PNS, resulting in fluctuations in the luminosity of the emitted \ns, and the emission of gravitational waves (\gw)
(see, e.g., \cite{Mirizzi:2015eza,Kuroda:2017trn,Andresen:2016pdt,bernhard14} and references therein).

Indications of SASI were first identified in two-dimensional (2D) numerical simulations \cite{Blondin:2002sm,Blondin:2006fx,Marek:2008qi,Marek:2007gr,bernhard14,Nakamura:2014caa,Summa:2015nyk}, and then confirmed by three-dimensional (3D) simulations as well \cite{Blondin:2006yw,Iwakami:2008qj,Fernandez:2010db,Hanke:2013jat,OConnor:2018tuw,Vartanyan:2019ssu,Walk:2019miz}. 
The precession frequency (and therefore the frequency of the \n\ modulations) was found to be between a few tens of Hz and 200 Hz \cite{Tamborra:2013laa,Kuroda:2017trn,Walk:2018gaw,Walk:2019miz} depending on the progenitor mass, nuclear equation of state (EOS), and progenitor rotation. A possible
correlation of the SASI-modulated neutrino and \gw\ signals has been studied in \cite{Kuroda:2017trn},
which also demonstrated that a \gw\ SASI signature could be contaminated by other effects (e.g., neutrino-driven convection and the associated turbulence). A multi-messenger analysis joint with \ns, which could clarify the presence of SASI in \gw, is particularly motivated. 

While the frequency of the SASI is expected to be mostly related to the mechanical properties of the PNS, the duration of SASI signatures in \ns\ and GW reflects the duration of the phase when the shock wave is stalled, before either being launched to drive an explosion, or dying out so the star implodes directly into a black hole (failed supernova). 
Indeed, progenitors at the interface of the successful and failed explosions tend to exhibit longer periods with SASI \cite{Walk:2019miz}.

At this moment,  SASI is a hypothesis -- supported by numerical simulations -- that awaits observational tests. Neutrinos and \gw\ are the only messengers that can, for a future galactic \sn, directly probe this phenomenon and provide measurements the relevant parameters (such as the SASI frequency and amplitude).   
Such measurements will clarify the properties of the PNS, the nuclear EOS, ultimately the yet-uncertain supernova explosion mechanism. The phase difference between GW and neutrino luminosity observed at Earth could also in principle (for an uncertainty-free signal at the source) probe propagation effects, like the time delay due to the \ns\ being massive \cite{Lund:2010kh,Beacom:1998yb}. It also carries the potential to estimate the different depths of the main production zone of neutrinos (the neutrinosphere) and of GW \cite{Kuroda:2017trn}. 

The theme of this paper is the detectability of SASI signatures in the neutrino luminosity as recorded at neutrino detectors on Earth, and the potential of estimation of its main phenomenological parameters. The SASI-induced modulation of neutrino emission has been studied previously on the base of both two-dimensional \cite{Rampp:2002bq,Buras:2005rp} and three-dimensional \cite{Hanke:2013jat,Mueller:2012sv} SASI-dominated supernova simulations. The neutrino signal in terms of its Fourier power spectrum was analyzed \cite{Lund:2010kh,Lund:2012vm,Tamborra:2013laa,Walk:2019miz,Migenda:2016xnc} in order to assess the detectability of SASI activity. The minimum requirement for signal detection was established by stating that the power spectrum of signal has to exceed the one of the background \cite{Lund:2010kh,Lund:2012vm,Tamborra:2013laa,Tamborra:2014aua}.

In this work we advance the topic to a more quantitative level, by establishing a framework which is new in the context of \n\ data analyses. This methodology is an implementation of the maximum likelihood principle, and uses the probability distribution of the observed power at different frequencies. As part of the likelihood-based analysis 
 we also address the question of parameter estimation, and  compare the results for the parameter variances to the optimally possible variance according to a Fisher matrix analysis of the problem. 

The present paper is intended as a first step towards a joint description of the problem for \ns\ and GW, which is left for future work.

The paper is structured as follows. In Sec. \ref{sec:general}, generalities are given on SASI and on \n\ detection. In Sec. \ref{sec:maxlik}, our methodology to establish the presence of SASI in a \n\ signal is presented, and results are shown using a specific numerical simulation as a test-bed of the method. Parameter estimation is then addressed in Sec. \ref{sec:parameter}, and a discussion follows in Sec. \ref{sec:disc}. Three appendices offer proofs and technical details to the interested reader. 

\section{Generalities}
\label{sec:general}

\subsection{Supernova neutrino detection}

We consider \n\ detections in two different experimental settings. The first is a water \ck\ detector at the Megaton mass scale, like the planned Hyper-Kamiokande (\hk\ from here on) \cite{Abe:2018uyc}.
For simplicity, only the main detection channel, inverse beta decay ($\barnue + p \rightarrow n + e^+$), is included here.  Individual positrons are detected via their \ck\ photon signature with high efficiency and excellent time resolution (microseconds or less 
\cite{Abe:2018uyc}). Therefore, here 
an ``event" from a \sn\ burst indicates an individual \n\ interacting within the volume of the detector. Background events due to other \n\ sources, cosmic rays or detector impurities -- which in principle could mimic \sn\ \ns\ events -- are negligible for a galactic \sn\ \cite{Scholberg:2012id}. 

 The number of \sn\ \n\ events in the detector is directly proportional to the number of target particles in the detector (and therefore to its mass), and it scales like the \n\ number flux, i.e., proportionally to $D^{-2}$, with $D$ being the distance to the star. 
  As a reference, here the expected \hk\ mass of 0.44 Mt and 100\% detector efficiency will be used; 
  results for different detector masses can thus be obtained by rescaling $D$. 
Given the microsecond recording time scale, the number of events $n_i$ in each millisecond time bin $[t_i , t_i+ 
\Delta t]$ is subject to Poisson statistical fluctuations (standard deviation $\sigma_i=\sqrt{n_i}$), with negligible correlations between different time bins.  

The second experimental setting refers to the kilometer-scale antarctic detector \ic\ \cite{Kopke:2017req}. 
There, the detection concept is designed for multi-TeV \ns, and is based on Digital Optical Modules (DOMs) positioned in geometrically sparse arrays in the antarctic ice. For a flux of  $\sim 10$ MeV \sn\ \ns, individual \n\ interactions (mostly from inverse beta decay, like in water) can not be resolved, however a surge of total photon count rate in the optical modules can be observed as a signal. In this context, an event is intended to be the observation of a photon in a DOM. 

In contrast with \hk, in \ic\ the background level is relatively high, at a rate of $\dot n \simeq 1340 ~{\rm ms^{-1}}$ \cite{Kowarik:2009qr,Lund:2010kh}. Therefore, the number events $n_i$ in each time bin is the number of photons recorded in the entire detector in that time bin, and is the sum of the contributions of the \sn\ signal (scaling like $D^{-2}$) and of background (fixed, and constant in time). Note that in this work we focus on the dominant emission signatures of anti-electron neutrinos (in both IceCube and Hyper-K), and we shall leave the consideration of multi-flavor interactions (albeit important, see an in-depth review by Mirizzi et al. \cite{Mirizzi:2015eza}) for future work


\subsection{SASI: physics and numerical predictions}
\begin{figure*}[htb]
	\centering
     \includegraphics[width=0.45\textwidth]{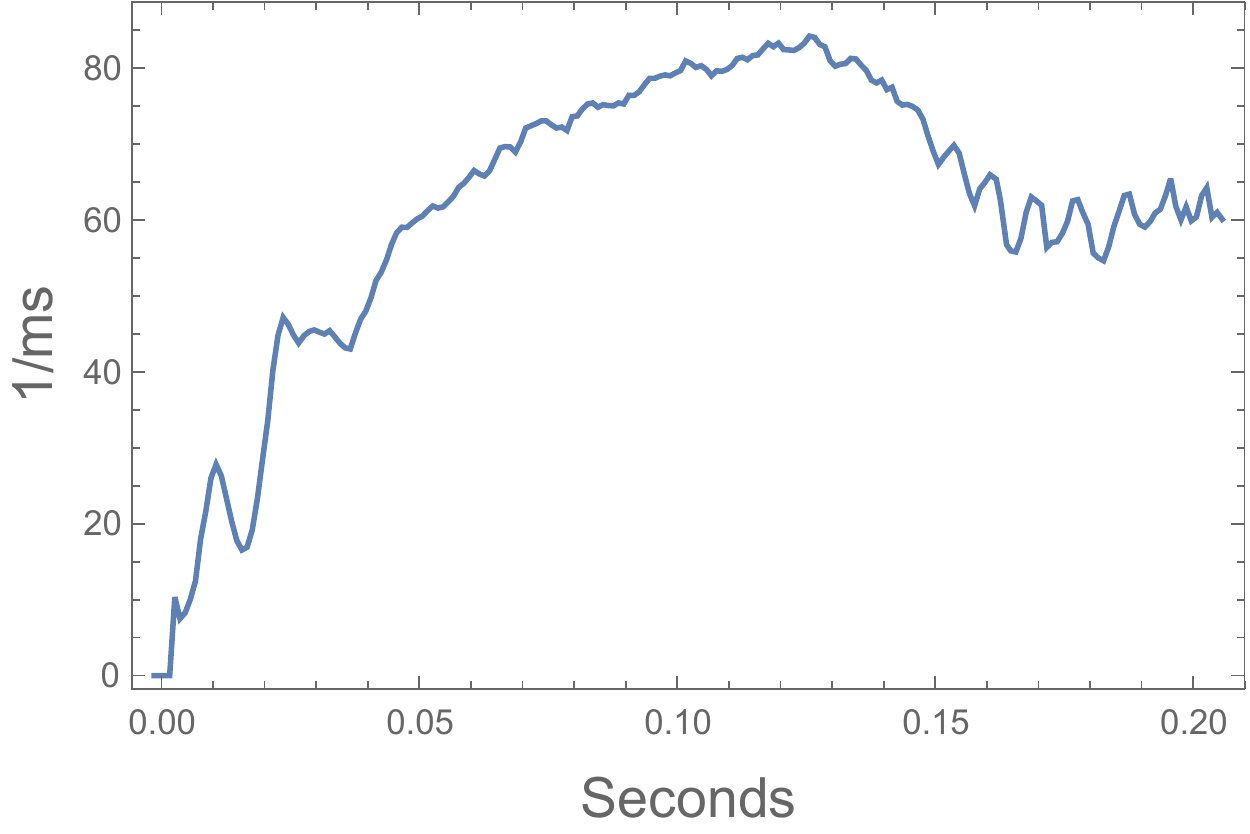}
	\caption{Predicted neutrino event rate  at \hk\ from the \kur\ model of a 15 \msun\ progenitor with the SFHx equation of state \cite{Kuroda:2017trn}, for a star at distance $D=10$ kpc. }
	\label{fig:knuratefull}
\end{figure*}
We use the numerically calculated \n\ event rates for \ic\ and \hk\  from a 3D general relativistic (GR) simulation (model SFHx, where SFHx indicates the equation of state by \cite{Steiner:2012rk}) by Kuroda, Kotake, Hayama and Takiwaki  (\kur\ from here on) \cite{Kuroda:2017trn} as a test bed of a realistic scenario where \si\  effects are present in the neutrino luminosity.
They are shown in Fig. \ref{fig:knuratefull}. For simplicity, the observer's direction is taken along the polar (e.g., the $z$) axis of the source 
as a fiducial case, where the flux-projection effects and the detection efficiencies for estimating the event rates are taken into account
 following \citet{Tamborra:2014aua}.

In the KKHT model, the 3D hydrodynamics evolution is self-consistently 
followed from the onset of 
core-collapse of a $15M_\odot$ star \cite{Woosley:1995ip}, through core bounce,
up to $\sim$ 350 ms after bounce. As consistent with the outcomes from recent 3D models (e.g., \cite{Hanke:2013jat,Andresen:2016pdt,Yakunin:2017tus}), the 
hydrodynamic evolution is characterized by the prompt convection phase shortly after bounce
($T_{\rm pb}\lesssim 20$ ms with $T_{\rm pb}$ the postbounce time),
 then the linear (or quiescent) phase ($20 \lesssim T_{\rm pb}\lesssim 140$ ms),
 which is followed by the non-linear phase when the vigorous activity of SASI was 
  observed for the model. The dominance of the SASI over neutrino-driven convection persists over $140 \lesssim T_{\rm pb} \lesssim 300$ ms, after which neutrino-driven convection dominates over the SASI 
  (see \cite{Kuroda:2016bjd} for more details).
  In \cite{Kuroda:2016bjd}, 
   the SASI frequency was roughly estimated as $\dot{M}/{M} \sim 100\, {\rm Hz}$,   
   
   where $M \sim 10^{-3} M_{\odot}$ and $\dot{M} \sim 0.1 M_{\odot}/{\rm s}$ denote the typical mass and mass accretion rate in the gain region, respectively, which is consistent with the numerically obtained SASI-modulated neutrino frequency (e.g., Figure 7 of \cite{Kuroda:2017trn}).

In the simulation, the Baumgarte-Shibata-Shapiro-Nakamura formalism was employed to evolve the metric 
\citep{Shibata:1995we,Baumgarte:1998te}, and the GR neutrino transport was solved 
by an energy-integrated M1 scheme \citep{Kuroda:2012nc}.  For simplicity, 
effects of neutrino flavor oscillations (e.g., the Mikheyev-Smirnov-Wolfenstein (MSW) effect \cite{Mikheev:1986gs}, and collective neutrino oscillations, see \cite{Mirizzi:2015eza} and references therein for a review) are neglected  in this study (see, e.g., \cite{Tamborra:2013laa} for a brief discussion of the validity of this approximation). 

Here the \sn\ burst  simulated by \kur\ will be used as representative of a future SASI-carrying  signal in the two detectors of interest. It will be compared with a similar signal that has no SASI features in it.   
Such null model is constructed by smoothing out the SASI oscillations from the original KKHT model. The smoothing is done by taking the event rates averaged over eight time bins, each  of 1 ms width, and performing a polynomial interpolation of these averaged rates. 
A zoomed-in plot of the \kur\ and smoothed out rates is given in Fig.  \ref{fig:compare} (black solid lines in left panes).

\section{Testing for SASI: likelihood ratio method}
\label{sec:maxlik}

In this section, we 
set up the formalism necessary to our statistical method.
Considering the oscillatory character of the SASI signatures, we  choose to work in the frequency space, and establish the discretized  power spectrum  of the \n\ time profile as the observable of interest.
The statistical behavior of the power spectrum is then presented. Finally, the likelihood ratio as test-statistics is defined and used to assess the detectability of the \si. 
We use the likelihood ratio as deciding statistic for the hypothesis test because of its optimality properties, which are described by the Neymann-Pearson Lemma \cite{2009fundamentals}.
 For clarity, in what follows the symbols with tilde (e.g., $\tilde N$) will indicate an actual outcome of a measurement, which is affected by statistical fluctuations. The same symbol without tilde (e.g., $N$) will be used for the mean, ``true" value of the same quantity.

\subsection{Neutrino time profile templates}
\label{sub:templates}

When data from a \sn\ burst are analyzed, it can be useful -- as it is often done in \n\ data analyses, see, e.g., \cite{Lund:2010kh} -- to fit the event rate time profile with simplified analytical templates that, while necessarily inaccurate, will allow to gain analytical understanding and to estimate the main phenomenological parameters. The latter can then be compared with predictions of detailed numerical simulations for greater insights into the microphysics at play. In this work, we use two parametric templates which characterize the main features of neutrino signals with and without the SASI activity respectively, to study the potential of a data analysis algorithm to identify the presence of \si. 

For the case with SASI activity
we choose a single frequency function:
\begin{equation}
R_2(t)=(A-n)(1+a \sin(2\pi  f_S t))+n~, 
\label{eq:mod2}
\end{equation}
where $A$ is the time-averaged event rate (the ``DC component") in the detector including instrumental noise (after possible experimental cuts), $a$ is the relative SASI amplitude, $n$ is the mean value of the background rate ($n=0$ for \hk),  and $f_S$ is the nominal frequency of the SASI. 
The second template,  for the case without \si, is a constant: 
\begin{equation}
R_{0}(t)=A~,
\label{eq:mod0}
\end{equation}
(with $A$ having the same meaning as in Eq. (\ref{eq:mod2})). 

In our method, only $f_S$ and $a$ will be treated as free parameters with respect to which the likelihood will be maximized. We assume that other relevant quantities, such as the DC component, $A$, and the starting time ($t_0$) and duration ($\tau$) of the \si\ activity, can be determined separately, by using theoretical priors, visual inspection, or a separate algorithm. 
For $A$, it is immediate to see that it can be  measured with high precision  (i.e., negligible uncertainty), without the need of a fit. Its relative uncertainty is $\delta A/A=1/\sqrt{N_{ev}}\ll 1$ where $N_{ev}$ is the total number of events, and $N_{ev}\gtrsim 2500$ in all the cases examined here  
(we also assume that systematic uncertainties on $n$ are negligible, because background rates can be measured precisely over years of data-taking).  

With regard to $t_0$ and $\tau$, here they are fixed to be  $t_0=155$ ms post-bounce, and $\tau= 55$ ms, consistently with the \kur\ simulation results (fig. \ref{fig:knuratefull}).
Fixing these quantities is legitimate in the spirit of answering the question whether there is indication of single-frequency fluctuations in a signal between two chosen (generic) instants of time. 
Realistically, in the context of a  more specific search for \si\  effects, $t_0$ and $\tau$ could be at first set using rough estimations from visual inspections of the data, in conjunction with expectations from the theory. Indeed, a delay in the onset of \si\ (relative to the bounce time) is expected considering that \si\ requires the shockwave to come to a stalling point. We checked that 3D numerical simulations roughly place $t_0$ in the interval $\sim 0.1-0.4$ s post-bounce \cite{Radice:2018usf,Powell:2018isq,Kuroda:2016bjd,OConnor:2018tuw,Muller:2011yi}, with $\tau$ being even more uncertain. It is possible that, by the time the next galactic \sn\ is observed, theoretical progress will be able to place stronger priors on $t_0$ and $\tau$. 
The method proposed here will be applicable to data with externally-estimated (not optimized) $t_0$ and $\tau$; the lack of optimization of these parameters will result in certain loss of power of the method, which can be overcome by generalizing the method to include $\tau$ and $t_0$ as fit parameters.

\subsection{Time series and power spectrum}
\label{sub:discretized}

Let us consider the events that are recorded in a detector after an initial time $t_0$, in time bins of width $\Delta=1~{\rm ms}$. The  $j$-th time bin then corresponds to the time $t_j=t_0+j\Delta$. 
The observed number of events in the same bin will then be $\tilde{N}(t_j)$, which is a random variable fluctuating around its mean $N(t_j)\simeq  R(t_j)\Delta$.

Following \cite{Lund:2010kh,Lund:2012vm}, we perform a discrete Fourier transform of the time series $\{\tilde N(t_j) \}$ over the time interval $[t_0,t_0+ \tau ]$, containing $N_{bins}=\tau/\Delta$ time bins. The discrete frequency resolution is then:
\begin{equation}
	\delta =\frac{1}{\tau}~,
	\label{eq:tau}
\end{equation}
which represents the minimum width of frequency bins for which statistical independence between adjacent bins can be realized (see the discussion in Appendix A). 
For our fiducial value $\tau= 55~{\rm ms}$,  the resolution is $\delta= 18~{\rm Hz}$~\footnote{Eq. (\ref{eq:tau}) implies that a that much longer SASI signature will result in more precise estimation of the frequency; this might be relevant for future work on long-stalling shockwaves.}. 
The Nyquist frequency becomes \cite{2009fundamentals}
\begin{equation}
	f_{\textit{Nyq}}=\frac{1}{2\Delta},
	\label{eq:fnyq}
\end{equation}
which corresponds to the frequency index
\begin{equation}
k_{\textit{Nyq}}=\frac{f_{\textit{Nyq}}}{\delta }=\frac{\tau}{2 \Delta}=\frac{1}{2}N_{bins}~.
\label{eq:knyq}
\end{equation}
 We define the discrete Fourier-transformed \n\ signal as:
\begin{equation}
\tilde h(k\delta )=\sum_{j=0}^{N_{bins}-1}\tilde{N}(t_j)e^{i2\pi j \Delta k\delta }~,
\label{eq:h}
\end{equation}
and the one-sided power spectrum, similarly to \cite{2009fundamentals} as:
\begin{equation}
\tilde P(k\delta )=\begin{cases}
2|\tilde h(k\delta )|^2/N_{bins}^2~~~ \text{for } 0<k\delta <f_{Nyq}~,\\
\\
|\tilde h(k\delta )|^2/N_{bins}^2 ~~~\text{for } k\delta =0 ~
\end{cases} 
\label{eq:power1}
\end{equation}
(here the identity $(|\tilde h(k\delta )|^2+|\tilde h(-k\delta )|^2)=2|\tilde h(k\delta )|$ was used).

The factor of $1/N^2_{bins}$ is included in order to fix the normalization, so that at $k=0$ we have $\tilde P(0)=(\tilde N_{ev}/N_{bins})^2$ (here $\tilde N_{ev}= \sum^{N_{bins}-1}_{j=0} \tilde{N}(t_j)$). 

\begin{figure*}[htp]
	\centering

	\includegraphics[width=0.45\textwidth]{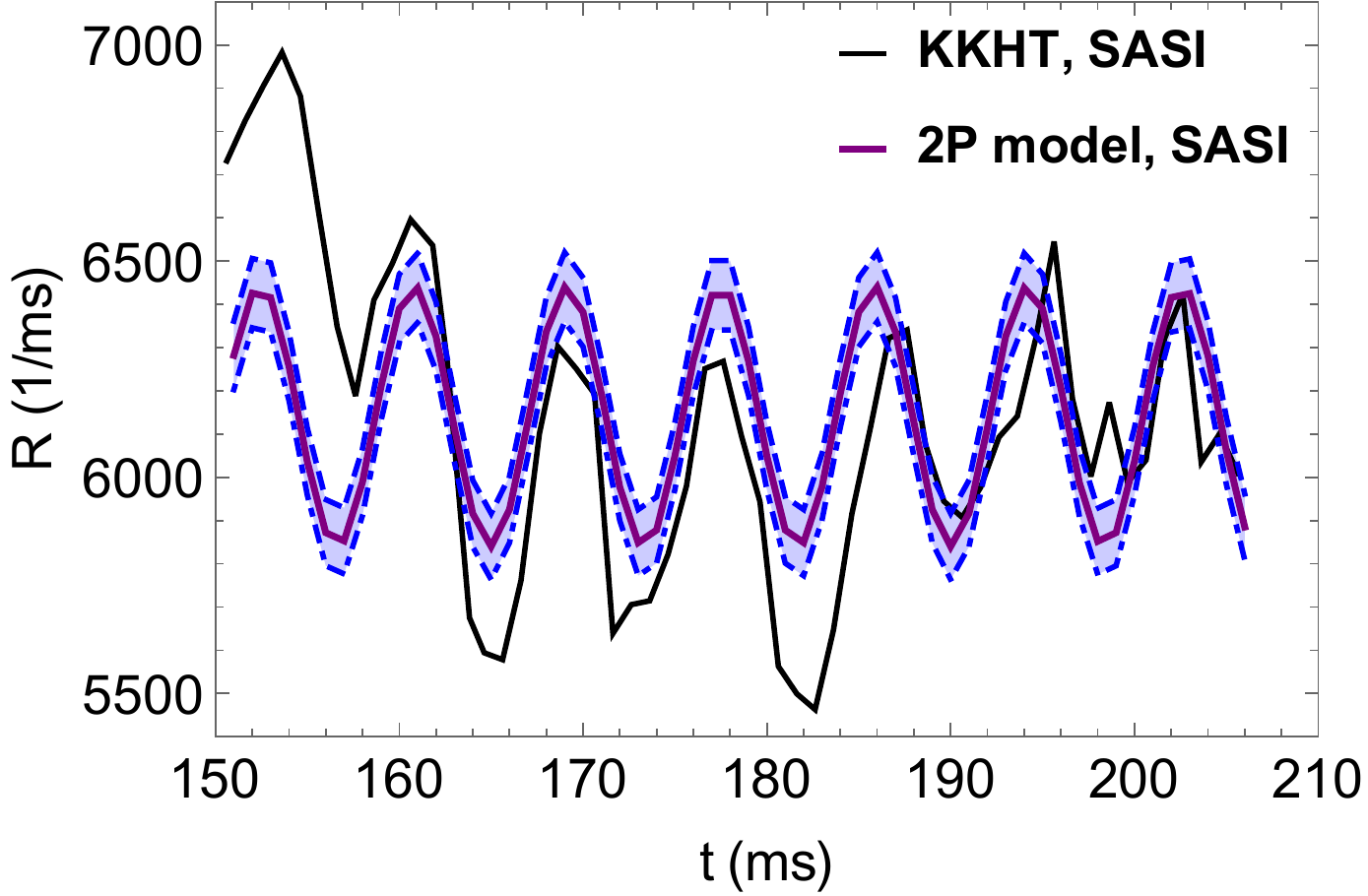}
	\includegraphics[width=0.45\textwidth]{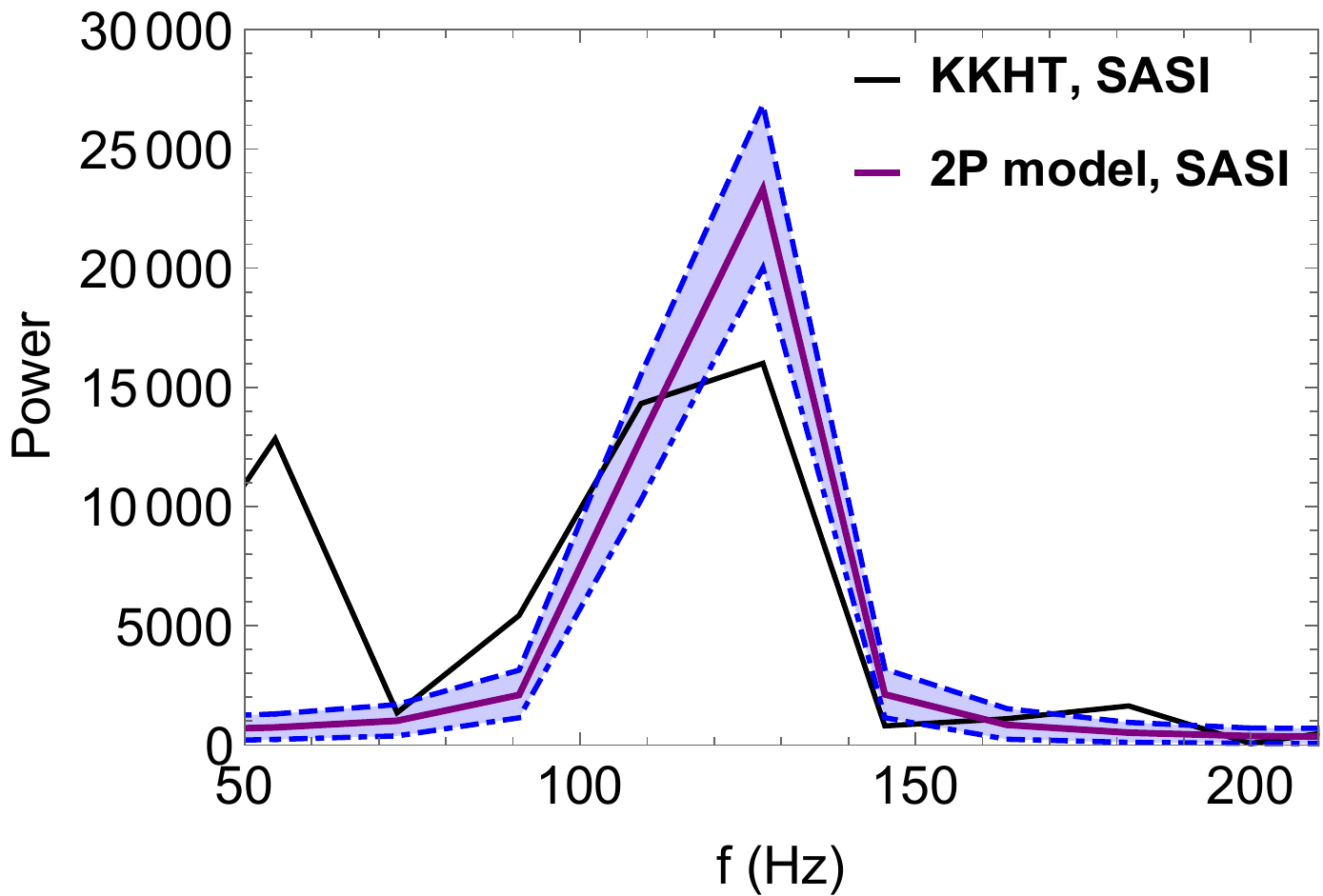}
	\includegraphics[width=0.45\textwidth]{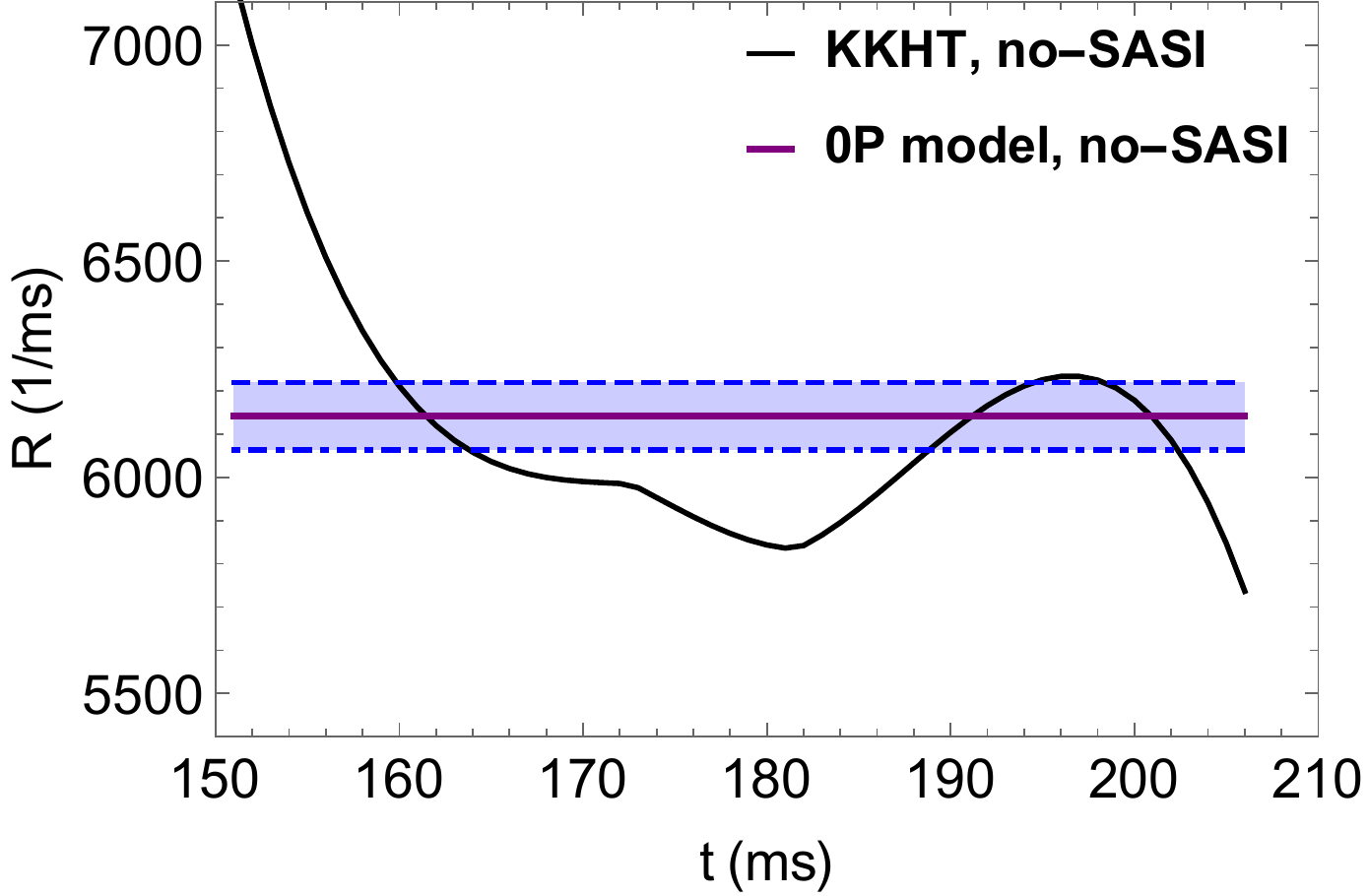}
	\includegraphics[width=0.45\textwidth]{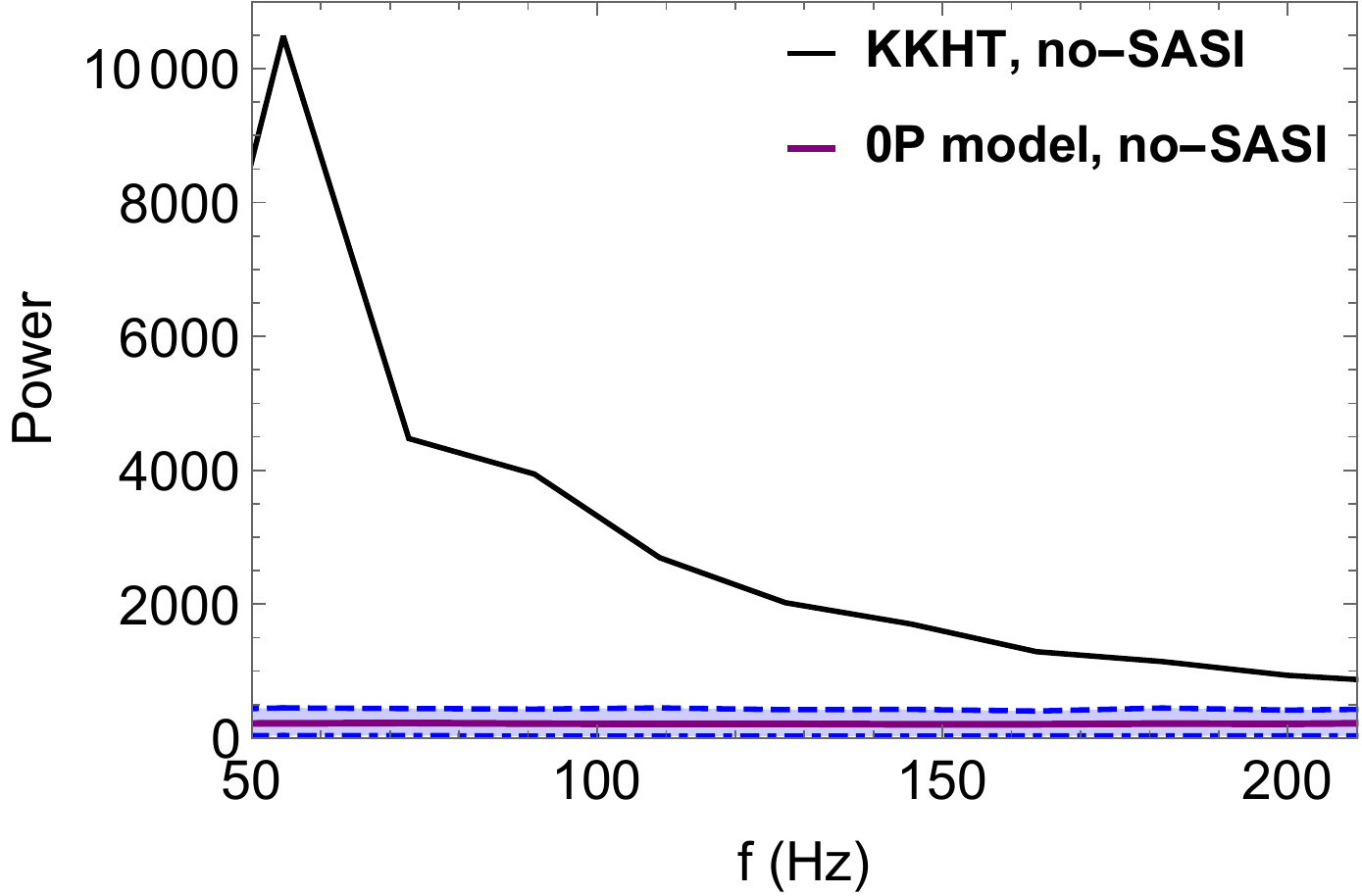}
	\caption{Neutrino event rate (left panels) and its power spectrum (right panels) at \hk\, for distance $D=1$ kpc.  Shown as solid black lines are a case where there is \si\ (upper panes, from the \kur\ model), and no SASI (lower panes, derived from the \kur\ model with smoothing, see text).  We also show (solid, purple curves) the predictions of the 2-parameter template (2P, Eq. (\ref{eq:mod2})) and of the 0-parameter template (0P, Eq. (\ref{eq:mod0})), for estimated best-fitting parameters
	($f_S=119.72$ Hz, $a=0.049$ and $A=6141.54$, see Eq. (\ref{eq:mod2}) and Table  \ref{tab:sasiPfa10}). The shaded (blue) bands characterize the probability density distributions with the width of one standard deviation. }
	\label{fig:compare}
\end{figure*}

Fig. \ref{fig:compare} shows an illustration of the discretized time profile, and the corresponding power spectra, for the \kur\ model, with and without SASI (as well as for the two templates in Eqs. (\ref{eq:mod2}) and (\ref{eq:mod0})).  For the latter, the parameters have been fit to maximize the likelihood (see eq. (\ref{eq:likeli}) in the following section) to best reproduce the general features of the \n\ event rates predicted by the \kur\ model. 
The figure shows that, qualitatively, the templates capture the main features of the realistic, numerically calculated time and frequency profiles. An exception is the peak at $f\sim 60$ Hz in the power spectrum of the no-\si\ model, which is not reproduced by the template. We checked that this peak is due to the ``wavy'' structure at $t\sim 180-200$ ms in the numerical model.

\subsection{The SASI-meter}
\label{sub:likelihoodR}

Let us now consider the series of power spectrum values at the discrete frequencies $k\delta$, $P(k\delta)$, and their statistical properties. 
Considering that (i) the probability that a single neutrino interacts in the detector is very small,  (ii) event counts in different time bins are statistically independent (see Sec. \ref{sec:general}), and (iii) $N(t_j) \gtrsim 10$ (large number approximation), we conclude that the binomial distribution for $N(t_j)$  approaches a Gaussian distribution with a variance proportional to the square root of the mean number (Poisson process): $s^2(t_j)=N(t_j)$. 
This implies (see the proofs in Appendices A and B) that the real part and imaginary part of the discrete Fourier transform, $h(k\delta)$ (Eq. (\ref{eq:h})), are also Gaussian-distributed,
and the probability distribution of the power spectrum $\tilde{P}$ at a given frequency is given by

\begin{equation}
\begin{split}
Prob(\tilde{P})&=\frac{N_{bins}^2}{4\sigma^2} \exp{ \left[ -\frac{N_{bins}^2}{4\sigma^2} \left(\tilde{P} + P \right)\right]}\\
&\times I_0\left( \frac{N_{bins}^2}{2\sigma^2} \sqrt{ \tilde{P}P} \right)~,
\label{eq:prob}
\end{split}
\end{equation}
where $I_0$ is the modified Bessel function of the first kind, and 
\begin{equation}
\begin{split}
\sigma^2=\frac{N_{ev}}{2}~.
\end{split}
\label{eq:sigma2}
\end{equation}

The object of this study is to perform a hypothesis test for the presence of \si. 
There is evidence from numerical simulations that the SASI only develops within a certain range of frequencies from a few tens of Hz to  about 250 Hz \cite{Mueller:2003fs,Murphy:2009dx,Yakunin:2010fn,Mueller:2012sv,Andresen:2016pdt,Kuroda:2016bjd}. Therefore, we apply a frequency cut, and restrict the analysis to the interval from 54 Hz to 216 Hz.  The corresponding range of wavenumbers is $k=3,4,5,...,12$. 
In addition to being motivated by estimates of the \si\ frequency, the cut is instrumental to exclude a large peak at low frequency due to the spectral leakage \cite{2009fundamentals} from $0$ Hz.

Let us now define the likelihood that a given observed power series vector, $\mathcal{\tilde P}= \{  \mathcal{\tilde P}_k \}$ (i.e., the series of powers for discrete wavenumbers $k$) is a realization of a certain hypothesis, which can be described by a parametric template.
It is defined as:  
\begin{equation}
L(\mathcal{\tilde P},\Omega)=\prod_{k=3}^{12}Prob(\mathcal{\tilde P}_k, P_k(\Omega))~,
\label{eq:likeli}
\end{equation}
where $P_k(\Omega)$ is the power predicted by the template, and $\Omega$ indicates the set of parameters of the template.

Given two hypotheses (i.e., two templates) with parameters $\Omega$ and $\Omega_0$, and a fixed observed set $\mathcal {\tilde P}$, the likelihood ratio is:
\begin{equation}
\mathcal{L}(\mathcal{\tilde P})=\frac{Max_{\Omega}[L(\mathcal{\tilde P},\Omega)]}{Max_{\Omega_0}[L(\mathcal{\tilde P},\Omega_0)]}~. 
\label{eq:likeR}
\end{equation}
In the numerator (denominator), the first (second) hypothesis is used and the likelihood is maximized with respect to the parameters $\Omega$ ($\Omega_0$).  
In this work, the templates in Eqs. (\ref{eq:mod2}) and  (\ref{eq:mod0}) will be used as representative of the \si\  and \nsi\ cases.
Their parameters are $\Omega=\{a, f_S \}$ and $\Omega_0=\{ Null\}$  respectively. 

It is intuitive to see how the likelihood ratio in Eq. (\ref{eq:likeR}) is sensitive to \si. Since our templates $R_2$ (Eq. (\ref{eq:mod2})) and $R_0$ (Eq. (\ref{eq:mod0})) capture well the main features of the neutrino event rates of the models with and without SASI respectively, as the \si\ features in the data become more pronounced, the numerator Eq. (\ref{eq:likeR}) is likely to increase (generally better fit for the $R_2$ template), while at the same time the denominator is likely to decrease (poorer fit for the $R_0$ template), so $\mathcal{L}$ is likely to increase. Vice-versa,  $\mathcal{L}$ will take lower values if the \si\ signatures in the data become weaker. 
Therefore, Eq. (\ref{eq:likeR}) serves as our \emph{``SASI-meter"} to identify the presence of SASI.

To assess the effectiveness of the \si-meter quantitatively, 
we need to find the probability distributions of $\mathcal{L}$ (or, equivalently, $\ln{\mathcal L}$) under the two hypotheses.  This was done by simulating (using a Monte Carlo method) $N_{st} = 10^3$ sets $  \mathcal{\tilde P} $  using the \kur\ model with and without \si, so we will have $\mathcal{L}_{S}\equiv \mathcal{L}(\mathcal{\tilde P}_{SASI})$ and $\mathcal{L}_{nS}\equiv \mathcal{L}(\mathcal{\tilde P}_{no-SASI})$,  and their probability density distributions, $Prob(\mathcal{L}_{S})\simeq  Prob(\mathcal{L}|S) $ (where $Prob(\mathcal{L}|S)$ indicates the ``true" probability distribution, which would be obtained in the limit $N_{st} \rightarrow \infty$) and $Prob(\mathcal{L}_{nS})\simeq Prob(\mathcal{L}|nS)$. 

A useful way to describe these two distributions, and compare them with one another, is to examine the probabilities that -- under the two hypotheses --   the likelihood ratio exceeds a certain threshold value, $\Lambda$: 

\begin{eqnarray}
&P_D=\int_{\mathcal{L}>\Lambda}Prob(\mathcal{L}|S) d\mathcal{L}\label{eq:pd}~,\\
&P_{FI}=\int_{\mathcal{L}>\Lambda}Prob(\mathcal{L}|nS) d\mathcal{L}~. \label{eq:pf}
\end{eqnarray}

$\Lambda$ usually represents a value of the likelihood ratio above which the \si\ hypothesis is accepted as true (``detection"). Therefore, $P_D$ takes the meaning of  SASI \emph{detection probability}, because it represents the probability that the method accepts the \si\ hypothesis as true  when the \si\ is in fact true. $P_{FI}$ then represents the \emph{false identification probability}, i.e., the probability that the \si\ hypothesis is accepted when in fact the \nsi\ hypothesis is the true one. 

The formalism discussed in this section becomes clearer in light of the results we have obtained, which are going to be illustrated next. 

\subsection{Results: SASI or no-SASI?}
\label{sub:results}

\begin{figure*}[htp]
	\centering
	\includegraphics[width=0.45\textwidth]{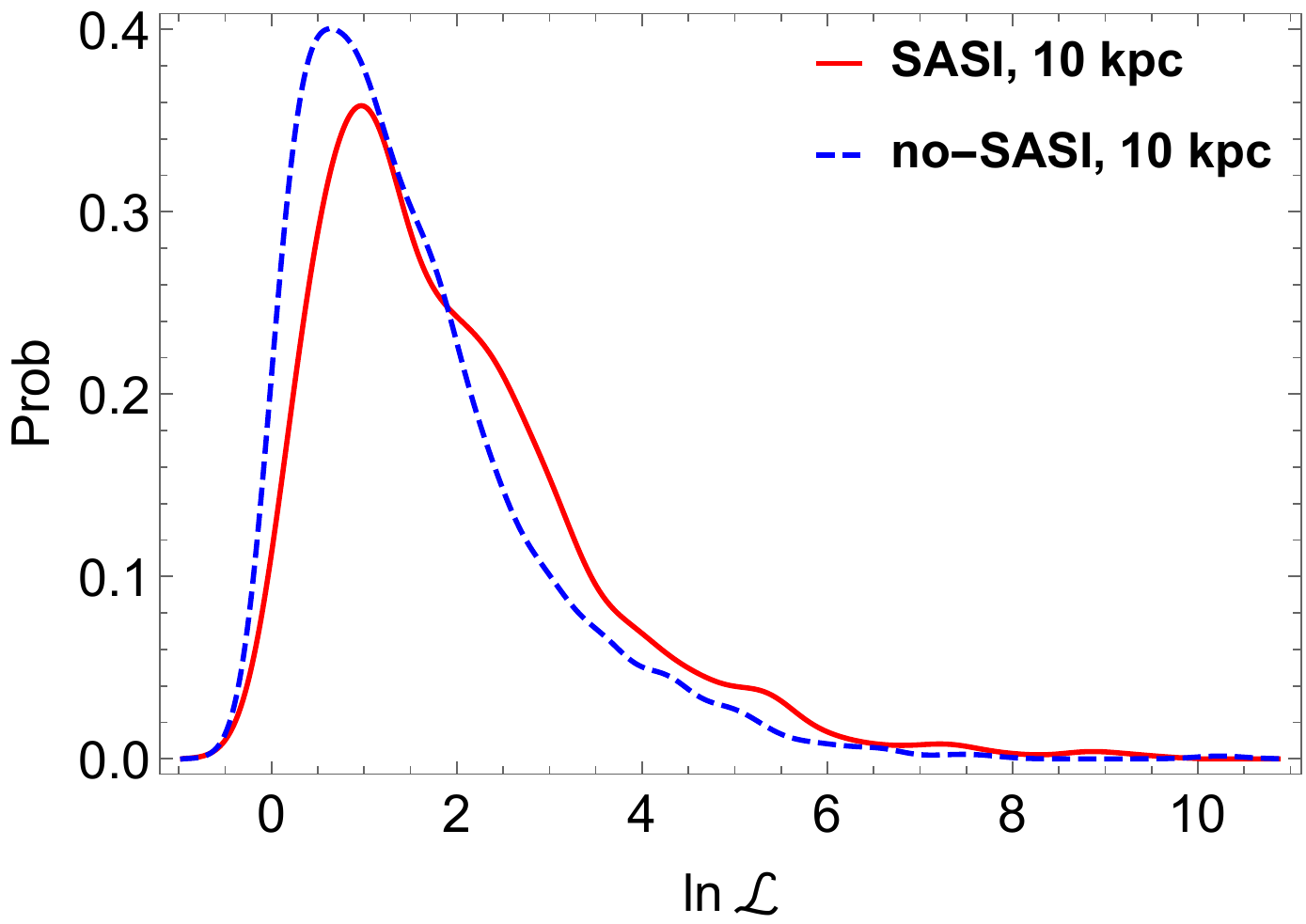}
	\includegraphics[width=0.45\textwidth]{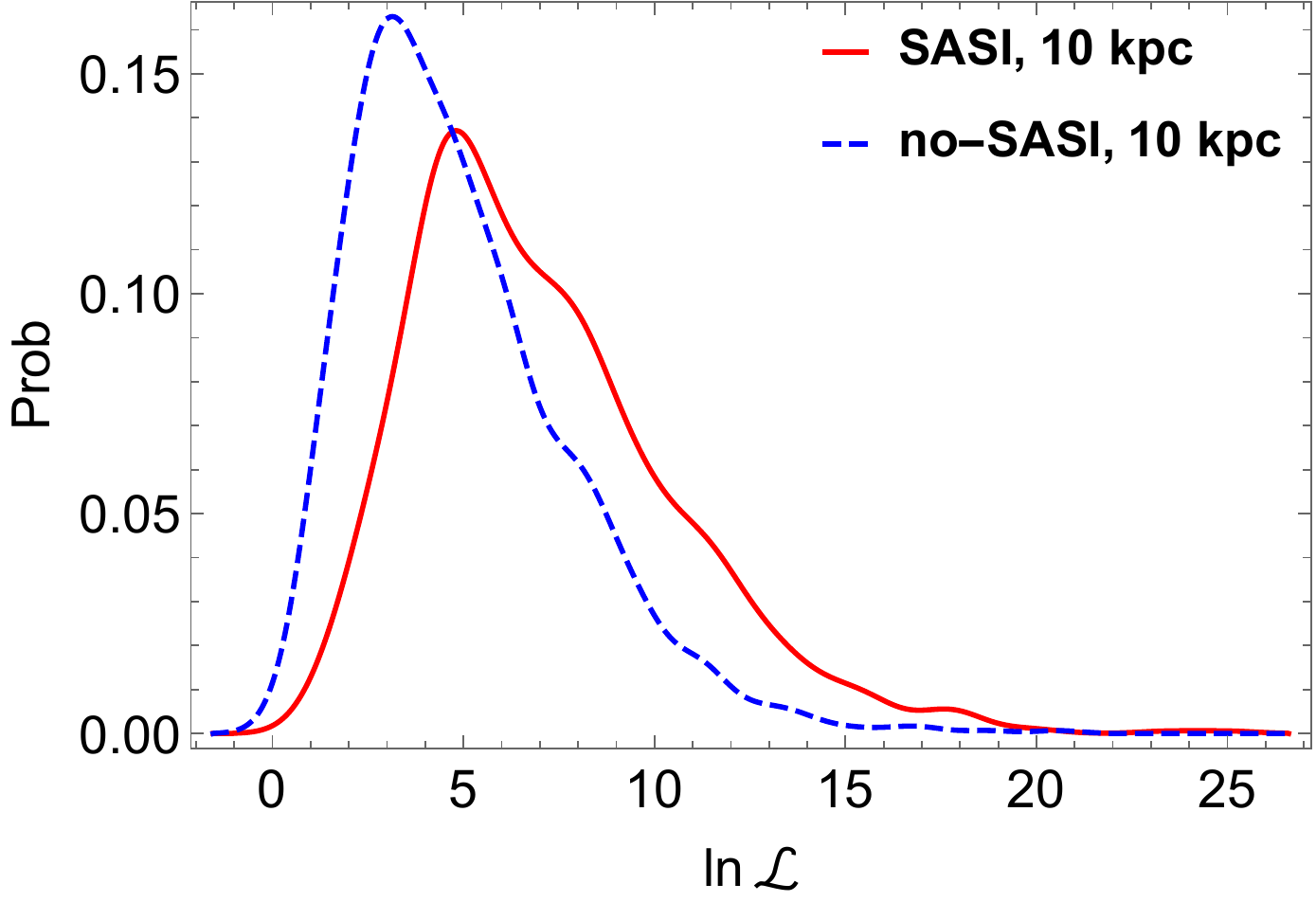}
	\includegraphics[width=0.45\textwidth]{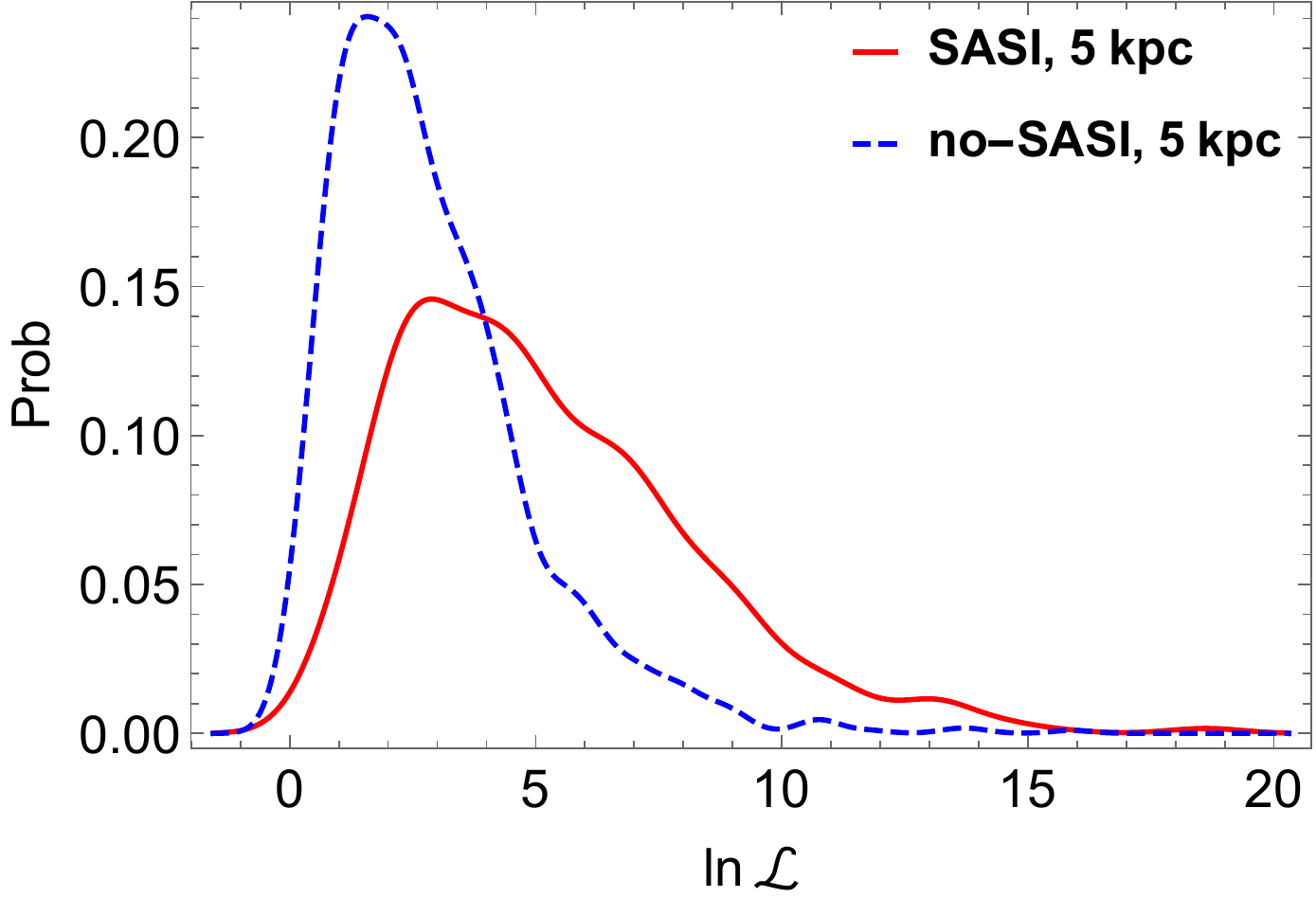}
	\includegraphics[width=0.45\textwidth]{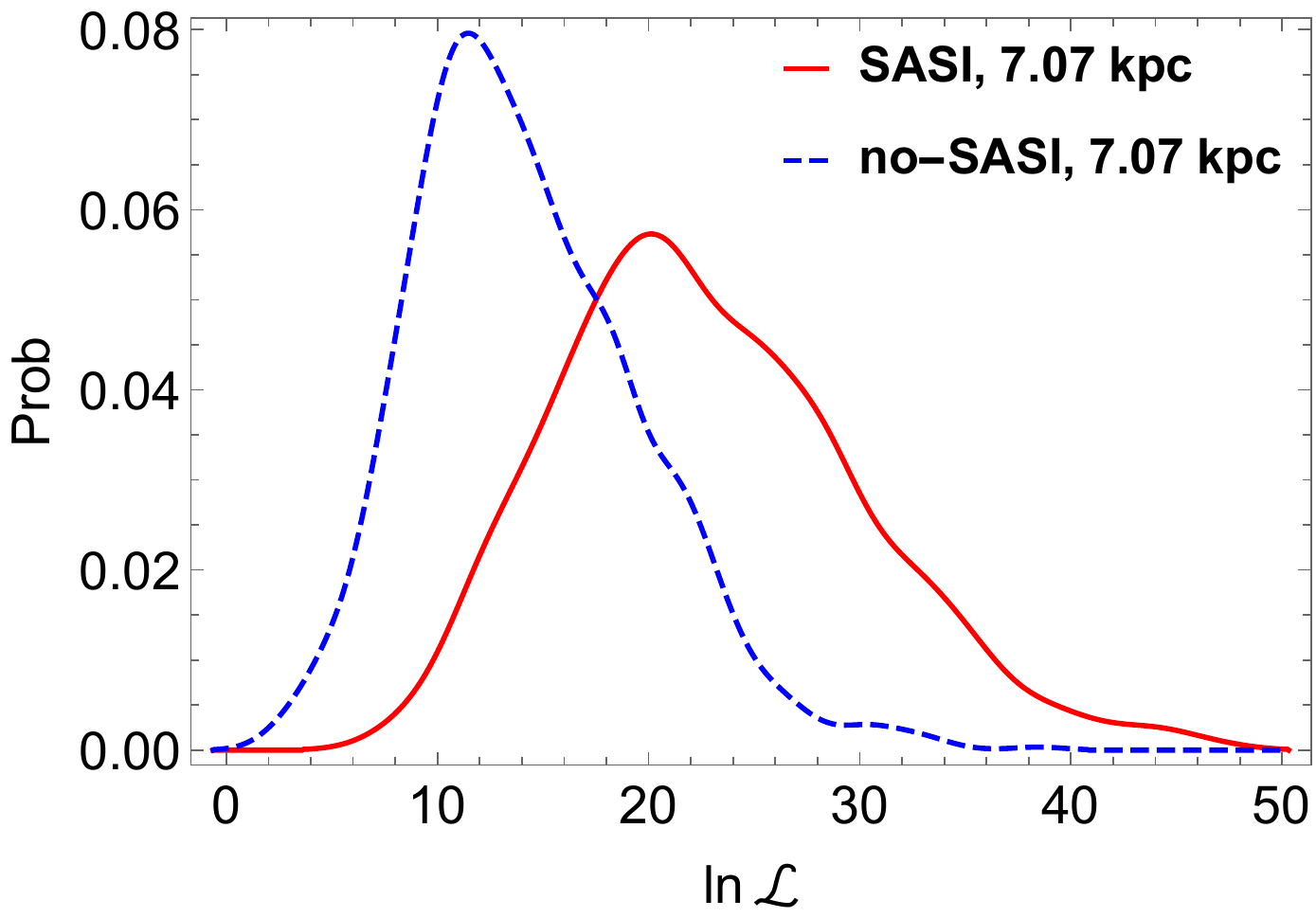} 
	\includegraphics[width=0.45\textwidth]{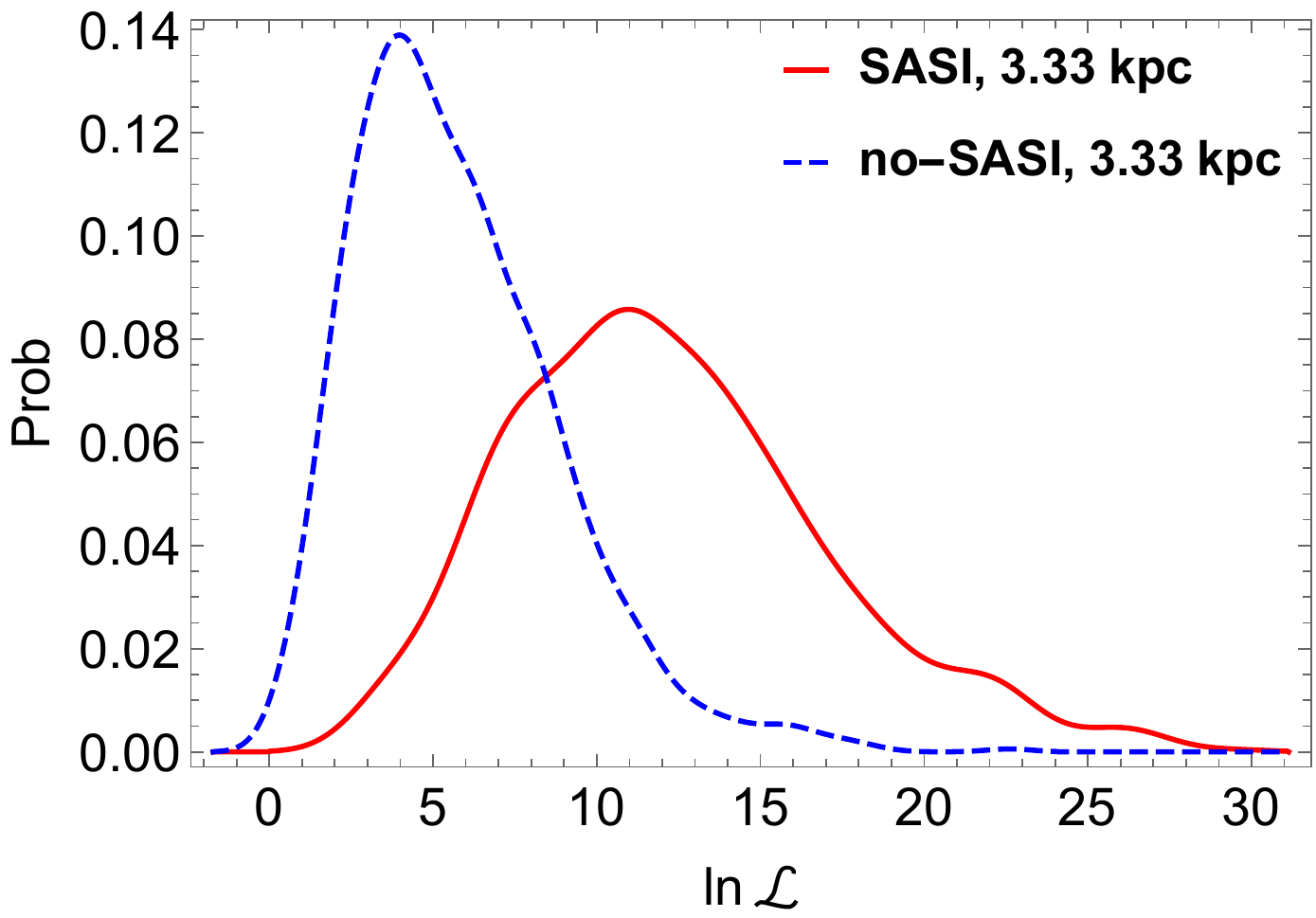} 
	\includegraphics[width=0.45\textwidth]{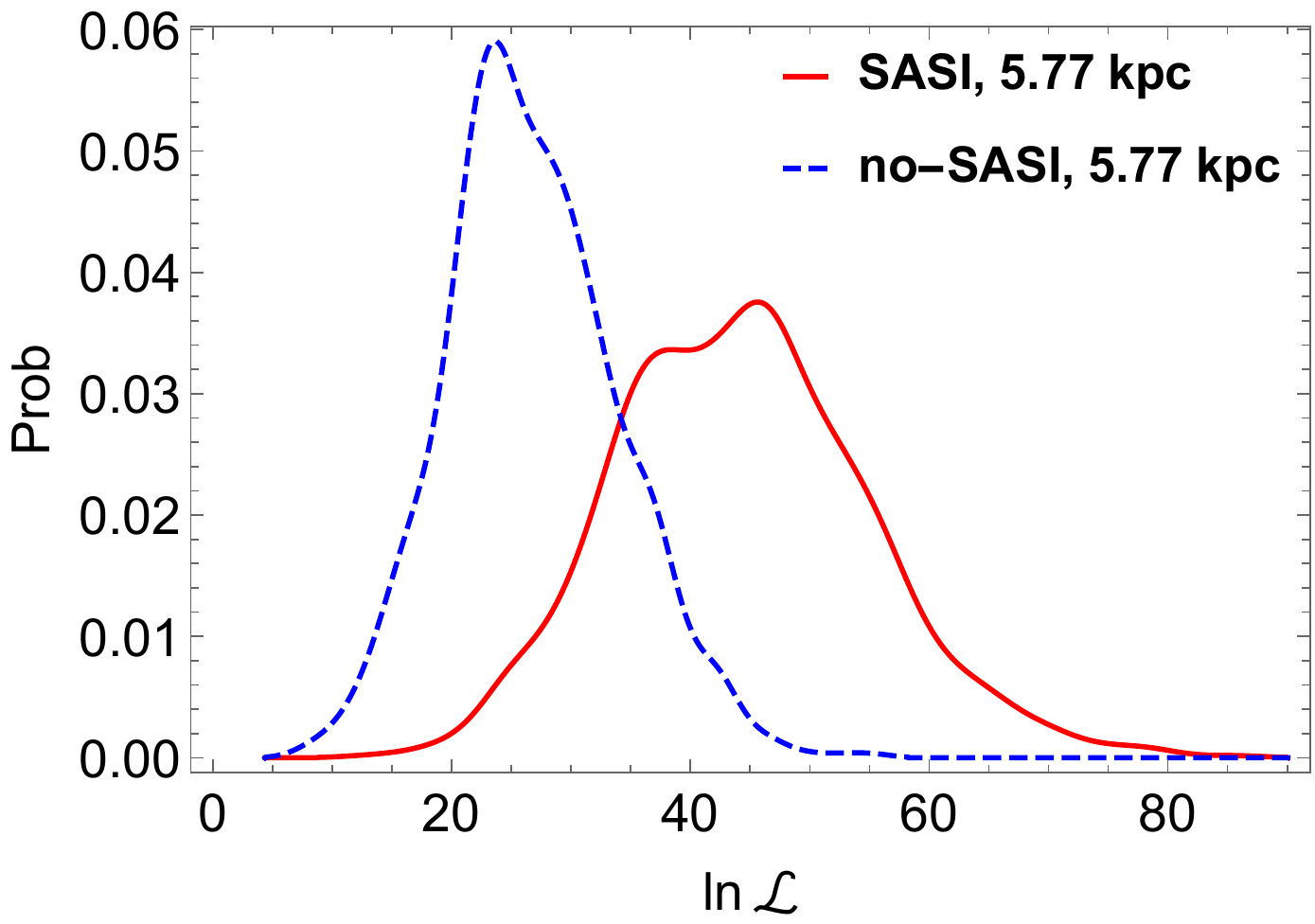}
	\includegraphics[width=0.45\textwidth]{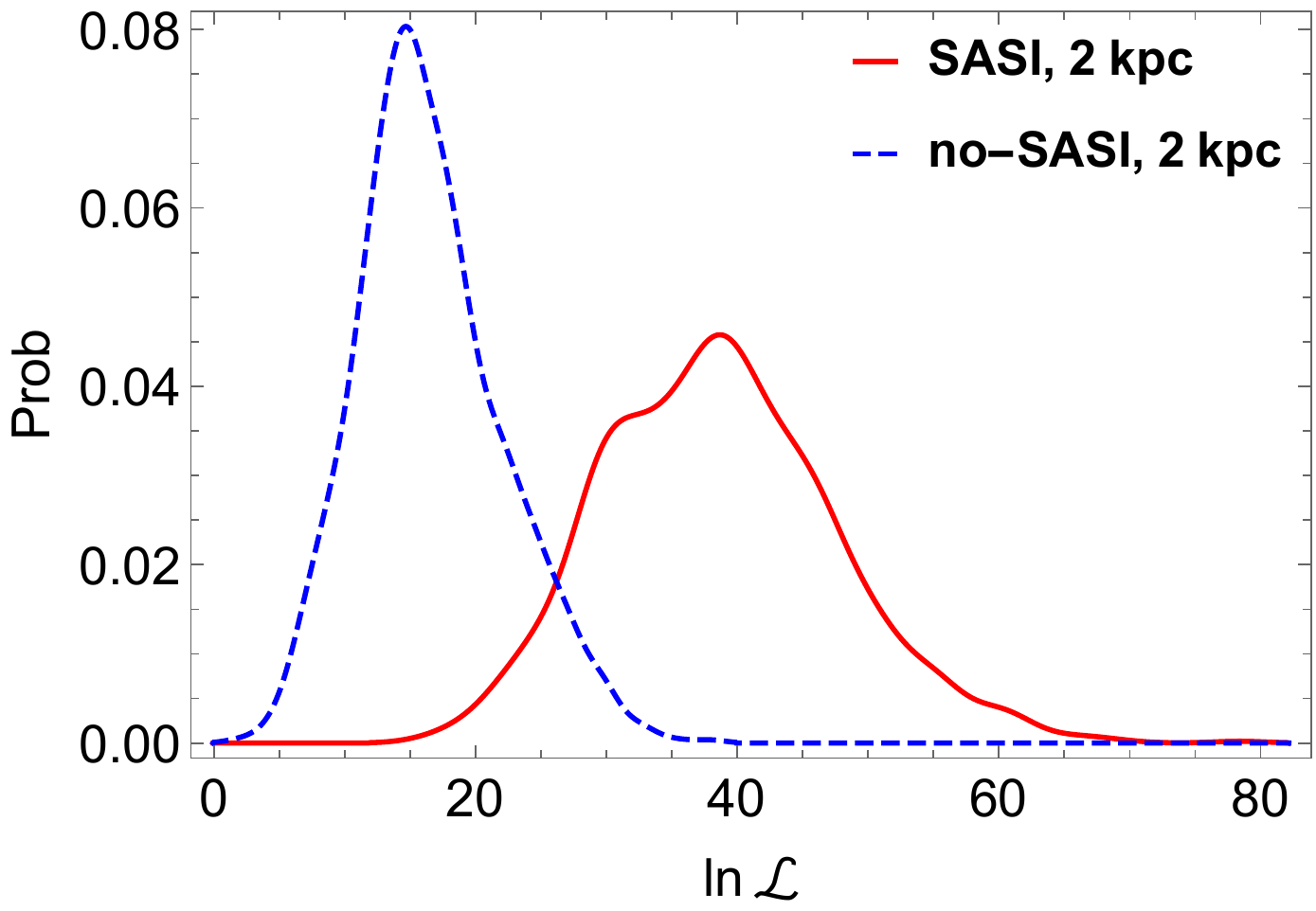} 
	\includegraphics[width=0.45\textwidth]{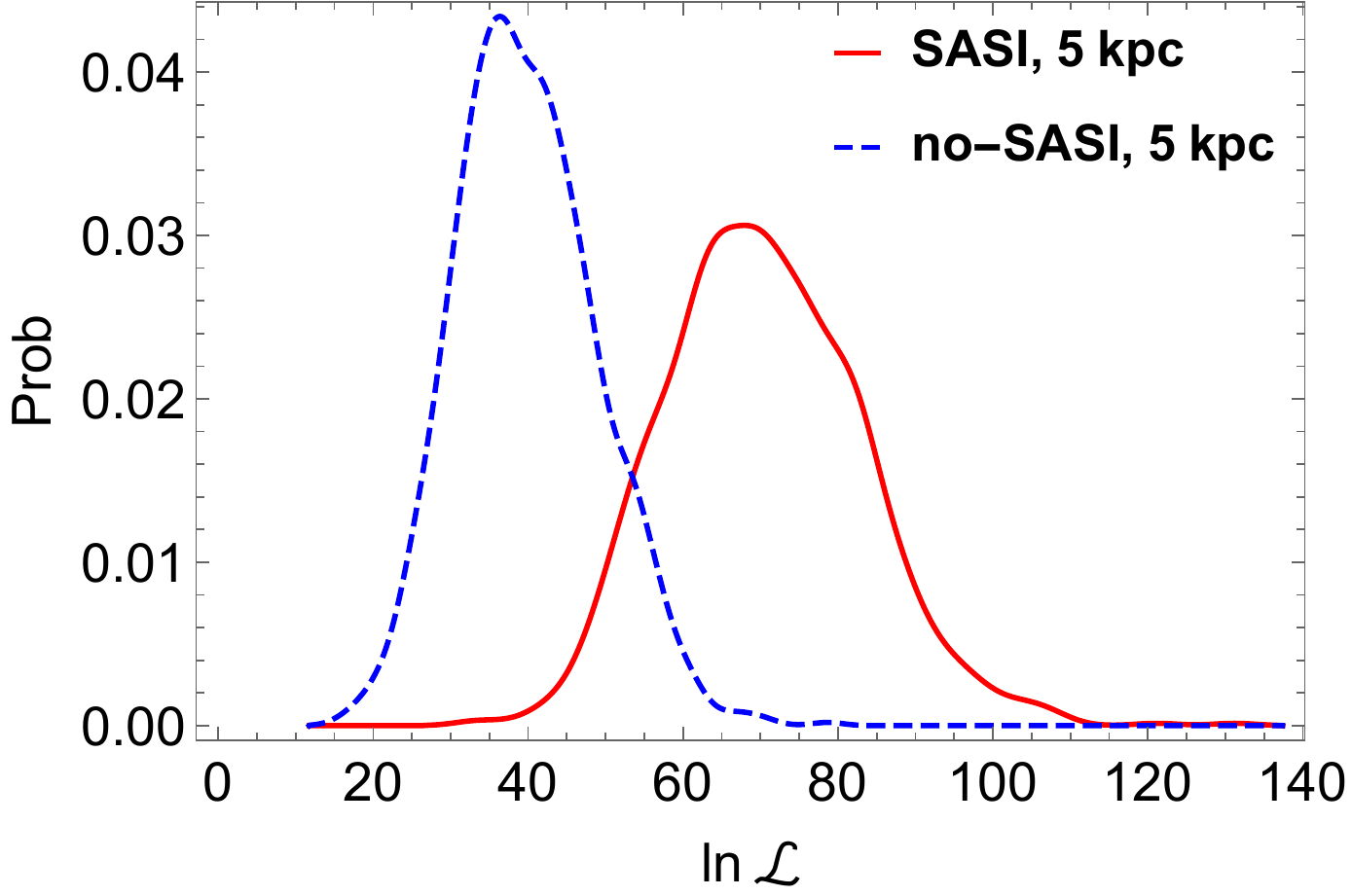} 
	\caption{Likelihood ratio probability distribution for SASI and no SASI case in \hk\ (left) and \ic\ (right), for different values of the distance $D$ to the star (chosen to correspond to integer increments of the number of events, see legends). The likelihoods have been obtained using simulated \n\ signal according to the KKHT model.} 
	\label{fig:likeR}
\end{figure*}

Our main results for  hypothesis testing are summarized in fig. \ref{fig:likeR}, for \hk\ and \ic, and for different distances to the \sn. For each detector and distance, the figure shows the probability distributions of $\ln \mathcal{L}_S$ and $\ln \mathcal{L}_{nS}$. 

We observe that, reflecting the expected sensitivity of our SASI-meter, for short distances the two distributions are widely separated, with the distribution for the \si\  (no-\si) case peaking at lower (higher) values of the likelihood ratio \footnote{Note that, all cases, the  logarithm of the likelihood ratio is positive, meaning that the two parameter template offers a better fit than the zero-parameters one. This is  simply a consequence of the larger number of parameters of one template with respect to the other. }. 
The separation means that, if the \si\ hypothesis is true, there is a large probability that the measured value of $\ln \mathcal{L}$ will fall in a region where the \nsi\ hypothesis is strongly disfavored (i.e., $Prob( \mathcal{L}|nS)\ll Prob( \mathcal{L}|S) $).   A similar argument holds if the \nsi\ hypothesis is true. We conclude, then, that for a relatively close supernova ($D \sim$ few kpc) the two hypotheses are likely to be distinguished with high confidence. 

The separation between the two probability distributions decreases as $D$ increases,
until, for $D \sim 10$ kpc, the \si\ and \nsi\ curves almost completely overlap, meaning that the two hypotheses are very unlikely to be distinguished. The dependence on the distance is due to how the size of the the statistical fluctuations increases with $D$, eventually overpowering the SASI, which therefore becomes invisible. 

\begin{figure}[htp]
	\centering
	\includegraphics[width=0.45\textwidth]{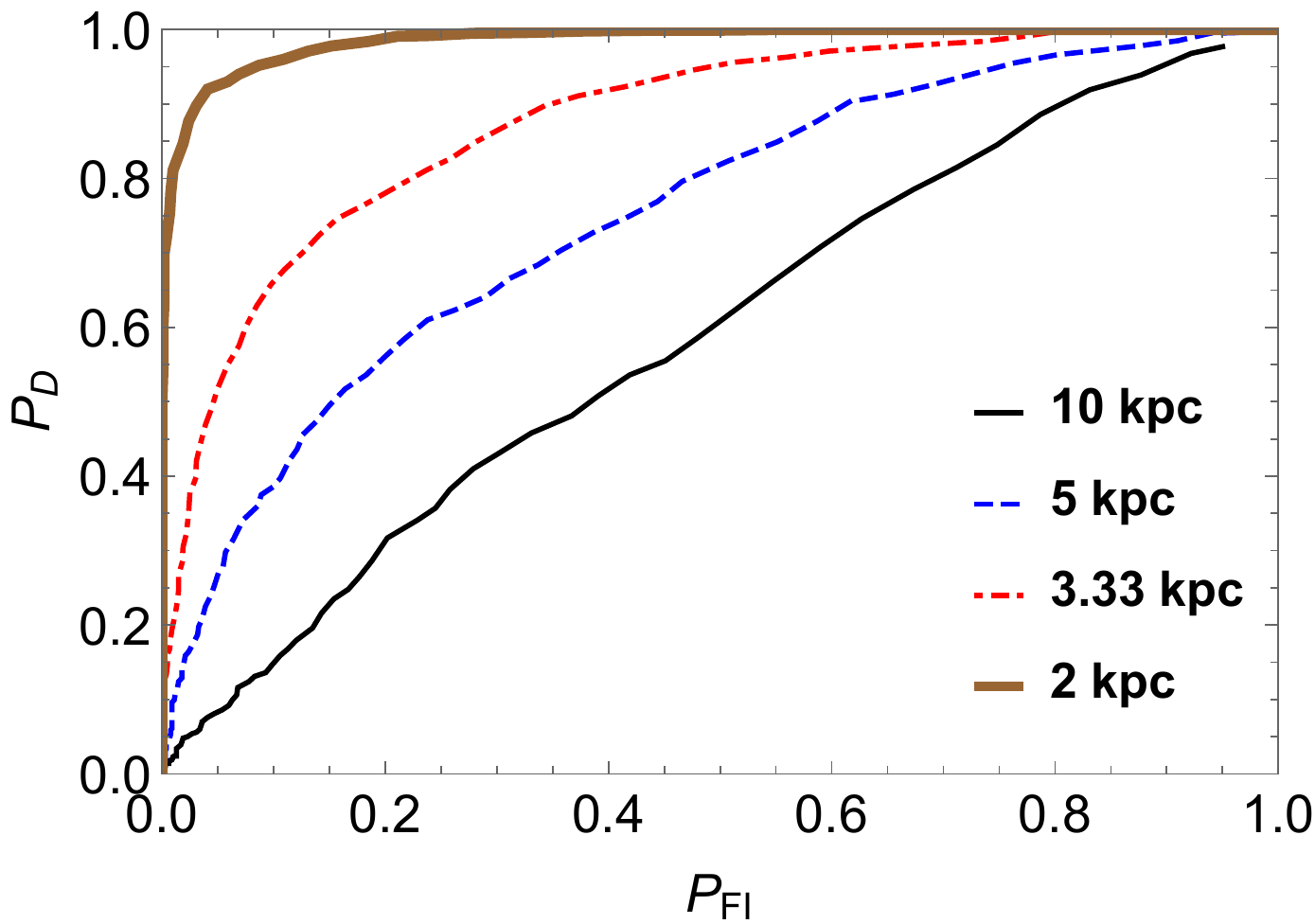}
	\includegraphics[width=0.45\textwidth]{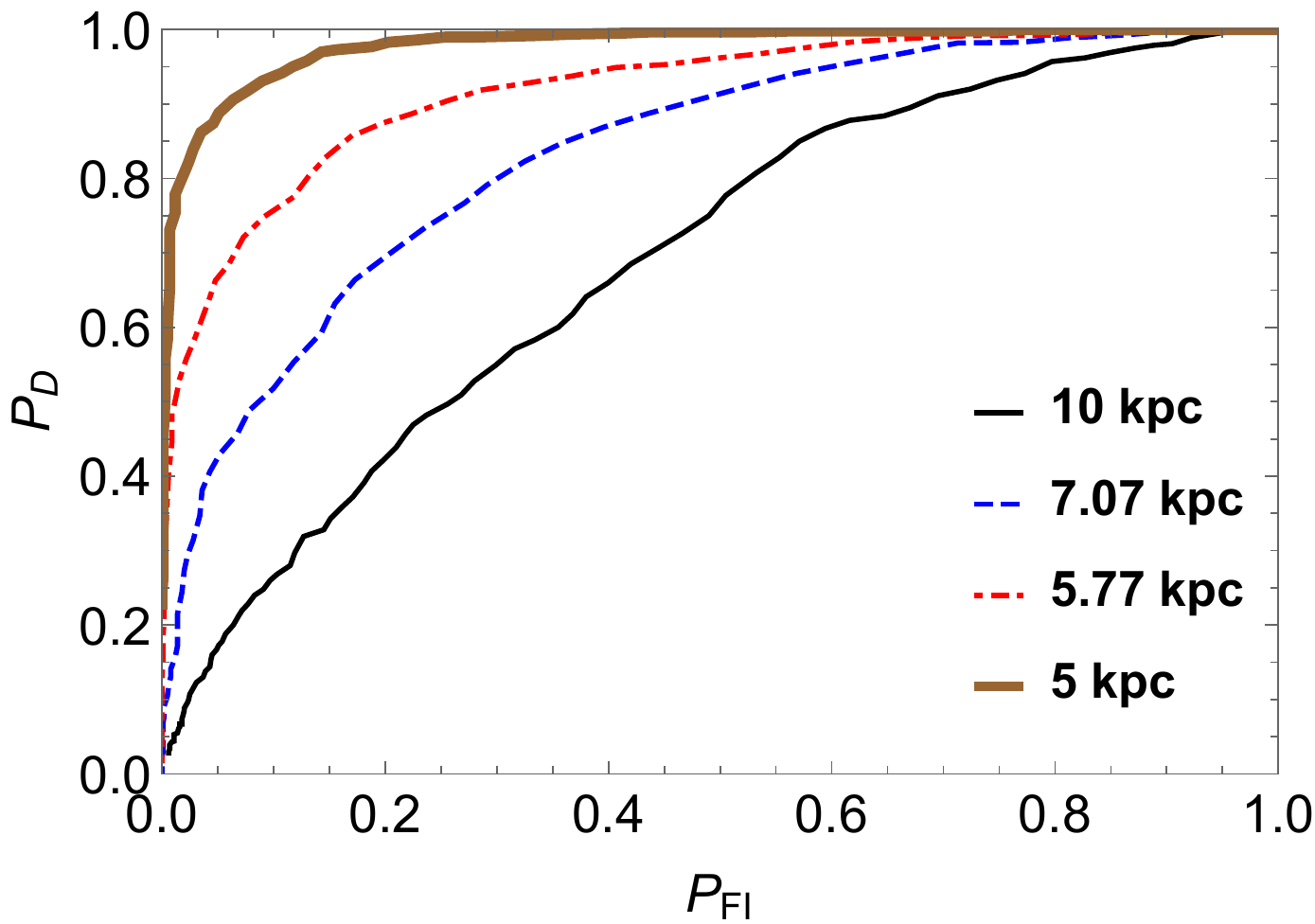} 
	
	\caption{Receiver operating characteristic curves based on KKHT model for \hk\ (top panel) and Ice Cube (bottom panel), for several distances to the supernova. See Eqs. (\ref{eq:pd})-(\ref{eq:pf}).}
	\label{fig:ROC}
\end{figure}

The trends shown in Fig. \ref{fig:likeR} are reflected in the behavior of the detection and false identification probabilities, $P_D$ and $P_{FI}$ (Eqs. (\ref{eq:pd}) and (\ref{eq:pf})). These are described by the Receiver Operating characteristic Curve (ROC).  The ROC is defined as the curve described in a plane by the points $(P_{FI}(\Lambda),P_D(\Lambda))$, where $\Lambda$ varies in the interval $[0,+\infty]$.  Fig. \ref{fig:ROC} shows the ROC for \hk\ and \ic\ for several distances from the star.  The plots show the general features of the ROC: it passes by the points $(0,0)$ and $(1,1)$ (corresponding to $\Lambda \rightarrow +\infty$ and $\Lambda \rightarrow 0$ respectively, see Eqs. (\ref{eq:pd}) and (\ref{eq:pf})). Furthermore, the curve lies in the region $P_D > P_{FI}$, as expected from Fig. \ref{fig:likeR}. A high detectability potential corresponds to a ROC where $P_D$ is as close as possible to $1$ and at the same time $P_{FI}$ is as close as possible to 0. For example, for \ic\ and $D=5$ kpc, the ROC passes by the point $(P_{FI},P_D)\simeq (0.1, 0.95)$, meaning that, if a 10\% false identification rate is considered acceptable, the likelihood ratio will establish the presence of the SASI in 95\% of the cases. The same situation is realized for \hk\ for $D\simeq 2$ kpc. Naturally, the ROC deteriorates as $D$ decreases, and ultimately (for $D \gtrsim 10$ kpc) it converges to the line $P_D=P_{FI}$, which corresponds to a \n\ signal with SASI being completely indistinguishable from a signal without SASI. The ROC curves allow to estimate the range where a fixed $P_D$ is achieved for a desired $P_{FI}$. If, e.g., we require the ROC to have $P_D \geq 0.7$ for $P_{FI}=0.1$, Fig. \ref{fig:ROC} indicates that the largest distance of sensitivity to the SASI is $D\simeq 6$ kpc for \ic\ and $D\simeq 3$ kpc for \hk.

\section{Parameter estimation}
\label{sec:parameter}

\subsection{Likelihood ratio and best fit parameters}

For the scenarios where the \si\ hypothesis is accepted as true (${\mathcal L}>\Lambda$), the next step is estimation of the  parameters. For definiteness, here we present results for $\Lambda$ that corresponds to $P_{FI}=0.1$ (Eq. (\ref{eq:pf})). 

In our method, the best fit values of the \si\ frequency, $\bar{f_{S}}$, and of the amplitude, $\bar{a}$,  are found as the values that maximize the likelihood $L(\tilde {\mathcal P},\Omega)$, within the process of constructing the likelihood ratio (Eq. (\ref{eq:likeR})).  From that process, we obtained the probability distributions  of $\bar{f_{S}}$, and $\bar{a}$. 
We then calculated the mean and standard deviation of $\bar{f_S}$ and $\bar{a}$. The standard deviation gives an estimate of approximately 68\% confidence level error with which an estimate of a given parameter can be obtained.

\begin{figure*}
	\includegraphics[width=0.45\textwidth]{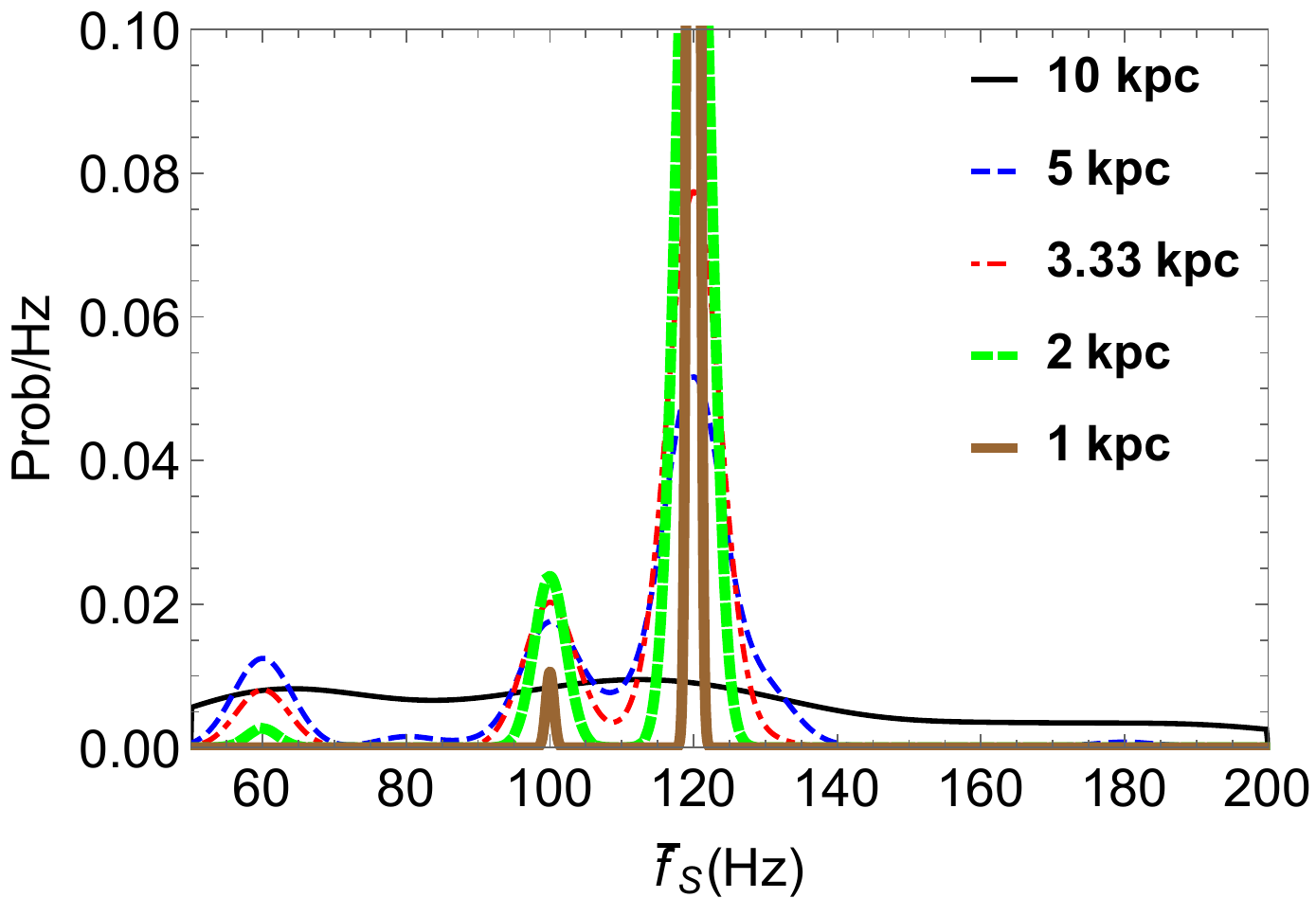}
	\includegraphics[width=0.45\textwidth]{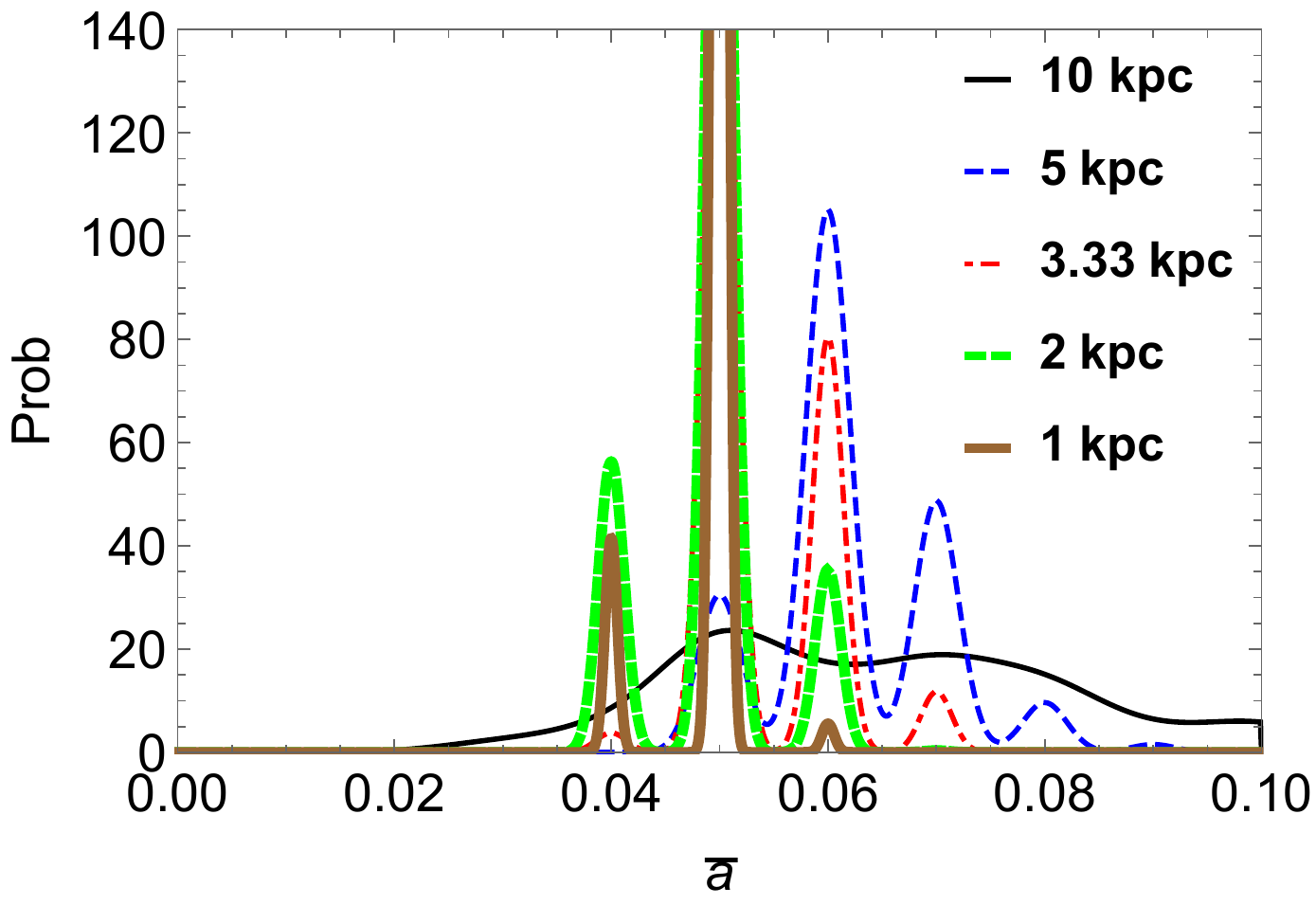}
 \includegraphics[width=0.45\textwidth]{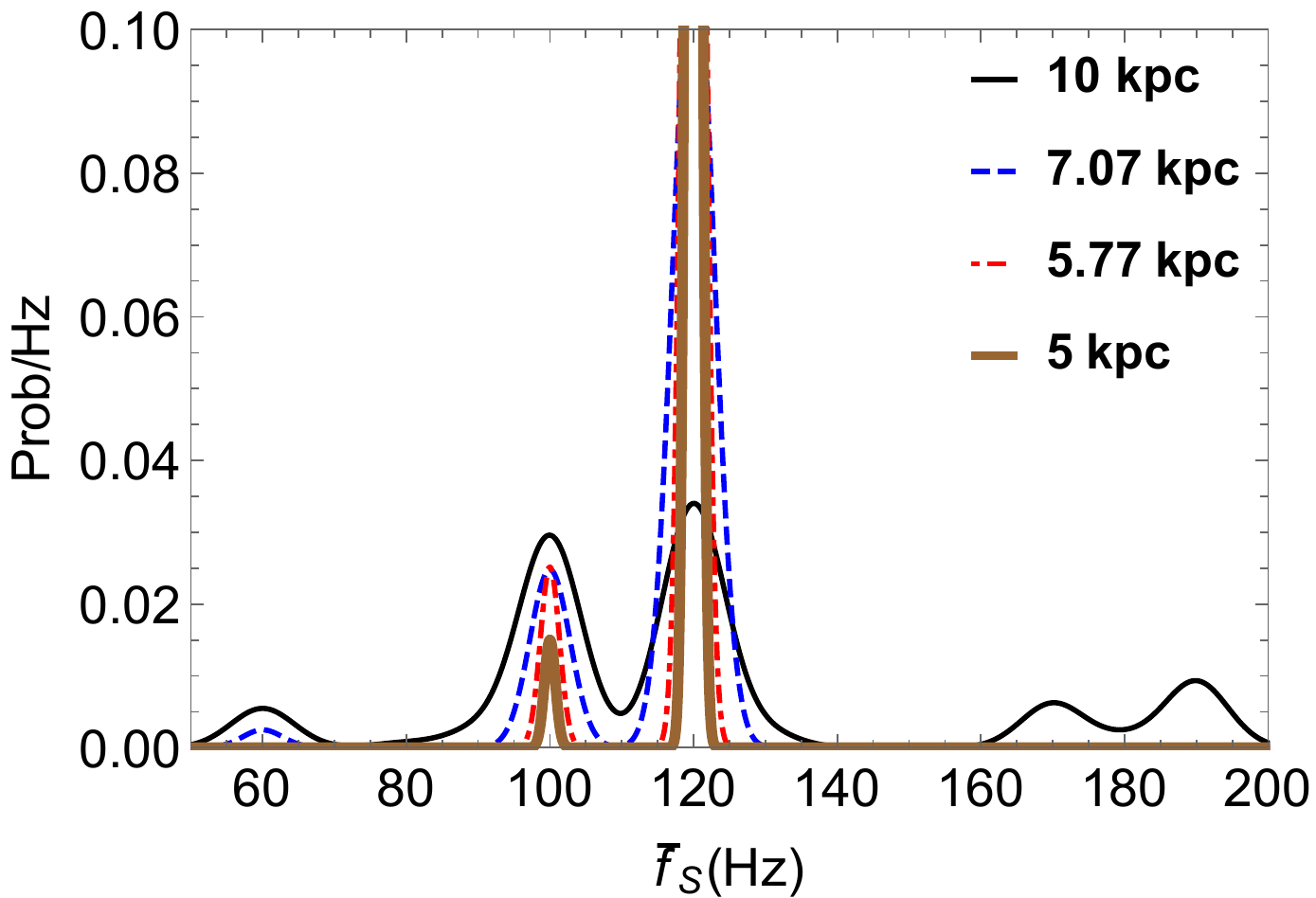}
	\includegraphics[width=0.45\textwidth]{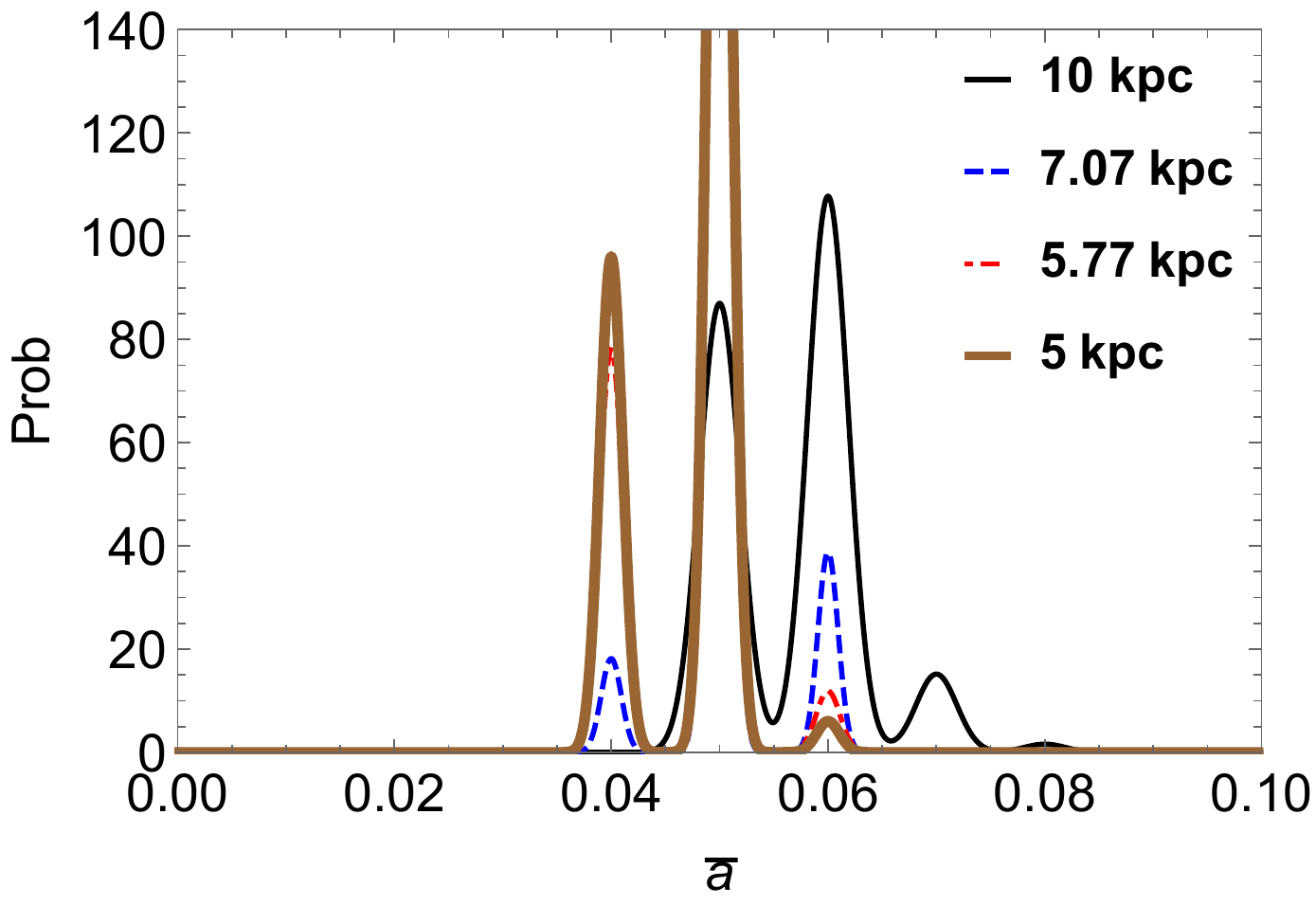}

		\caption{Probability distribution of the best-fit values of the \si\ frequency, $\bar f_S$ (left), and oscillation amplitude, $\bar a$ (right), for Hyper-K (top) and for \ic\ (bottom), for several distances to the \sn. Only cases with sufficient statistical indication of \si\ activity are considered here, by imposing a threshold on the likelihood ratio (corresponding to $P_{FI}>0.1$, see text). Here, the likelihoods have been obtained using simulated \n\ signal according to the KKHT model.} 
	\label{fig:parameterHypIceC}
\end{figure*}

The results are shown in Fig. \ref{fig:parameterHypIceC}  and Tables \ref{tab:sasiPfa10} (for \hk) and \ref{tab:sasiIcePfa10} (for \ic).
For \hk\ and $D=10$ kpc, where the sensitivity to the \si\ is poor, the distribution for $\bar f_S$ is very broad, with roughly all values being equally probable. This indicates that, although there might be indication of an oscillatory behavior in the data (such that the likelihood ratio is above the threshold), such outcome is most likely to be due to random statistical fluctuations and not to \si.  An estimate of the frequency would have a large error and might not be physically meaningful.  The corresponding distribution for $\bar a$ is similarly broad for $\bar a\gtrsim 0.03$, indicating that, as long as there is indication of an oscillatory pattern in the data, its amplitude can vary widely, and is probably driven by statistical fluctuations. 

As $D$ decreases ($D \lesssim 5$ kpc or so) the distributions of both $\bar f_S$ and $\bar a$ start to concentrate around the physical values of the injected \si\ model, $\bar f_S \sim 120$ Hz and $\bar a \sim 0.05$, indicating a sensitivity to the physical \si\ signal above statistical fluctuations. This trend appears in Table \ref{tab:sasiPfa10} as well, where one can see the decrease of the standard deviation with the decreasing distance. We note that the width of the distributions for $a$ and $f_S$ depend in part on how the time structure of the \n\ signal in the \kur\ model is only roughly reproduced by the simplified template, Eq. (\ref{eq:mod2}). As a consistency test, we checked that using simulated data drawn from the simplified template has the (expected) effect of producing narrower parameter distributions \footnote{For very high statistics signals, an accurate evaluation of the parameter distributions will  require a finer sampling of the parameter space than done here. We checked that the effect of the finite sampling is minor in our results. }.  

We caution the reader about the meaning of the multiple peaks that appear in the distributions in Fig. \ref{fig:parameterHypIceC}: these peaks reflect the discrete structure of the power spectrum series $\{\tilde P_k \}$ which is being analyzed, which has a resolution (frequency bin size) of about 20 Hz (see eq. (\ref{eq:tau}) and Fig. \ref{fig:compare}), and therefore do not have a direct physical meaning.

The probability distributions and tabulated values (Table \ref{tab:sasiIcePfa10}) for \ic\ show a  structure and dependence on $D$ similar to those for \hk.  A difference is that at $D=10$ kpc, the sensitivity to \si\ is not completely washed out by the statistical fluctuations, so it might be possible to obtain a (coarse) measurement of $f_S$.    

\subsection{Fisher Information Matrix and minimum uncertainties}

In this section we aim at comparing the standard deviations of the \si\ parameters obtained using the likelihood ratio method with the theoretical lower bound in the accuracy. The latter is given by the \cmr\ lower bound \cite{2009fundamentals}, and is derived from the Fisher Information Matrix (\fm). 
We  begin  by  summarizing  the main formulae of the FIM formalism; these will then be applied to the case at hand. 

\begin{table*}
	\caption{\label{tab:sasiPfa10}
	Mean and standard deviation of the parameter distributions in Fig. \ref{fig:parameterHypIceC}, for \hk.  The numbers in the parentheses are the \cmr\ lower bounds, calculated using Fisher matrix in the time domain. For larger $D$, where the selection effects on ${\mathcal L}$ are strong, a direct comparison is not meaningful and therefore the \cmr\ bounds are not shown. 
	}
	\begin{ruledtabular}
		\begin{tabular}{cccccc}
			SASI &$10$ kpc &$5$ kpc&
			$3.33$ kpc  &$2 $ kpc &$1$ kpc\\
			\hline
			f(Hz)& 111.77 & 109.2 & 111.67 &116.49
			& 119.72 \\
			$\delta f$(Hz)& 42.31 & 22.61& 16.7 &9.64(0.08)
			&2.35(0.04) \\
			a& 0.065 & 0.062& 0.054 &0.049
			& 0.049  \\
			$\delta a$& 0.017& 0.008 & 0.0059 &0.0053(0.0047)
			& 0.0026(0.0023)  \\
			
		\end{tabular}
	\end{ruledtabular}
	
\end{table*}

\begin{table*}
	\caption{\label{tab:sasiIcePfa10} 
Mean and standard deviation of the parameter distributions in Fig. \ref{fig:parameterHypIceC}, for \ic. The numbers in the parentheses are the \cmr\ lower bounds, calculated using Fisher matrix in the time domain. For larger $D$, where the selection effects on ${\mathcal L}$ are strong, a direct comparison is not meaningful and therefore the \cmr\ bounds are not shown. }  
	\begin{ruledtabular}
		\begin{tabular}{ccccc}
			SASI&$10 $ kpc &$7.07$ kpc &$5.77$ kpc &$5 $ kpc \\
			\hline
			f(Hz)& 120.45 &115.57&118.33& 119.43\\
			$\delta f$(Hz)& 33.36  &10.85 &5.53 (0.078)& 3.34 (0.063)\\
			a& 0.057 &0.050&0.048& 0.048\\
			$\delta a$& 0.0064 &0.0036&0.0047(0.0041)& 0.0046 (0.0034) \\
		\end{tabular}
	\end{ruledtabular}
\end{table*}

 Let us consider a generic template $R(t_i)$ for the event rate at discrete times, $t_i$ ($i=1, 2,..., N$), which depends on a set of parameters, $\theta_\alpha$ ($\alpha=1,2,3,..,K$) (note that, for our choice of unitary bin size, $\Delta=1$ ms, the event rate and the number of events are numerically the same. Here we omit the factor $\Delta$ to keep the notation compact). The \fm\ is a $K \times K$ matrix, found from the probability distribution. We define the joint probability as:
\begin{equation}
Prob(\tilde{R})=\prod_{i=0}^{N}Prob(\tilde{R}_{i}),
\label{}
\end{equation}
where $\tilde{R}$ is the series of observed   neutrino rate $\{\tilde{R}(t_1), \tilde{R}(t_2),...\tilde{R}(t_N)\}$ in time domain. The FIM describes how much each parameter affects the distribution via its second derivatives:
\begin{equation}
\Gamma_{\alpha \beta} =  \langle -\frac{\partial^{2} \ln{Prob(\vec{R})}}{\partial \theta_{\alpha} \partial \theta_{\beta}} \rangle~ ,
\label{eq:gamma}
\end{equation}

In the assumption that $Prob(\tilde{R}(t_{i}))$ is a Multivariate Gaussian Distribution in the time domain, the \fm\ reduces to the following expression (see Appendix C):
\begin{equation}
\Gamma_{\alpha \beta} = \mu_{\alpha}^{T} \Sigma^{-1} \mu_{\beta} + \frac{1}{2} Tr[\Tilde{c}_{\alpha} \Tilde{c}_{\beta}]~,
\label{eq:gamma2}
\end{equation}
where $\mu_{\alpha}^{T}$ and $\mu_{\beta}$ are  N-dimensional vectors (one component for each value of $t_i$), defined as: 
\begin{equation}
\mu_{\alpha} =  \frac{\partial \tilde{R}}{\partial \theta_{\alpha}}~,
\label{eq:mualpha}
\end{equation}
and
$\Sigma^{-1}$ is  the inverse of the $N\times N$ diagonal covariance matrix:
\begin{equation}
\Sigma^{-1} =
\begin{bmatrix}
{R(t_{1})}^{-1} & 0 & 0 & 0 & 0 \\
0 & {R(t_{2})}^{-1} & 0 & 0 & 0 \\
0 & 0 & {R(t_{3})}^{-1} & 0 & 0 \\
0 & 0 & 0 & \ddots & 0 \\
0 & 0 & 0 & 0 & {R(t_{N})}^{-1} \\
\end{bmatrix}~.
\label{cmatrix}
\end{equation}
Finally $\Tilde{c}_{\alpha}$ is defined as the inverse of the covariance matrix times the partial derivative of the matrix:
\begin{equation}
\Tilde{c}_{\alpha} = \Sigma^{-1} \frac{\partial \Sigma}{\partial \theta_{\alpha}}~.
\label{tildealpha}
\end{equation}

The \cmr\ bound on a parameter $\theta_\alpha$ is given by:
\begin{equation}
\delta \theta_{\alpha} \geq \sqrt{(\Gamma^{-1})_{\alpha \alpha}}
\label{eq:CRbound}
\end{equation}

We can now specialize the \fm\ formalism to our case, where the template is the one in Eq. (\ref{eq:mod2}), and we have two parameters, $\theta_1=a$ and $\theta_2=f_S$. Therefore: 
\begin{equation}
R(t_{i}) =R_2(t_i)=(A-n)(1+a \sin{(2 \pi f_{s} t_{i})}) + n
\end{equation}
\begin{equation}
\mu_{1} = (A-n)\sin{(2 \pi f_{s} t_{i})}
\end{equation}
\begin{equation}
\mu_{2} = 2 \pi t_{i} (A-n) a \cos{(2 \pi f_{s} t_{i})}~.
\end{equation}

The elements of Fisher matrix in time domain can be written analytically as below: 
\begin{equation}
\begin{aligned}
	\Gamma_{11}&=\sum^{N}_{i=1} \frac{(A-n)^{2} \sin{(2 \pi f_{s} t_{i}))}^{2}}{(A-n)(1+a \sin{(2 \pi f_{s} t_{i})}) + n}
	\\&+
	\frac{(A-n)^{2} \sin{(2 \pi f_{s} t_{i}))}^{2}}{2((A-n)(1+a \sin{(2 \pi f_{s} t_{i})}) + n)^{2}},
\end{aligned}
\label{eq:gamma11}
\end{equation}

\begin{equation}
\begin{aligned}
\Gamma_{12}&=\sum^{N}_{i=1} \frac{2a (A-n)^{2} \pi t_{i} \sin{(2 \pi f_{s} t_{i})} \cos{(2 \pi f_{s} t_{i})}}{(A-n)(1+a \sin{(2 \pi f_{s} t_{i})}) + n}
\\&+
\frac{a (A-n)^{2} \pi t_{i} \sin{(2 \pi f_{s} t_{i})} \cos{(2 \pi f_{s} t_{i})}}{((A-n)(1+a \sin{(2 \pi f_{s} t_{i})}) + n)^{2}},
\end{aligned}
\label{eq:gamma12}
\end{equation}

\begin{equation}
\begin{aligned}
\Gamma_{21}&=\sum^{N}_{i=1} \frac{2a (A-n)^{2} \pi t_{i} \sin{(2 \pi f_{s} t_{i})} \cos{(2 \pi f_{s} t_{i})}}{(A-n)(1+a \sin{(2 \pi f_{s} t_{i})}) + n}
\\&+
\frac{a (A-n)^{2} \pi t_{i} \sin{(2 \pi f_{s} t_{i})} \cos{(2 \pi f_{s} t_{i})}}{((A-n)(1+a \sin{(2 \pi f_{s} t_{i})}) + n)^{2}},
\end{aligned}
\label{eq:gamma21}
\end{equation}

and \begin{equation}
\begin{aligned}
\Gamma_{22}&=\sum_{i=1}^{N} \frac{4 a^{2} (A-n)^{2} \pi^{2} t_{i}^{2} \cos{(2 \pi f_{s} t_{i})^{2}}}{(A-n)(1+a \sin{(2 \pi f_{s} t_{i})}) + n}
\\&+
\frac{2 a^{2} (A-n)^{2} \pi^{2} t_{i}^{2} \cos{(2 \pi f_{s} t_{i})^{2}}}{((A-n)(1+a \sin{(2 \pi f_{s} t_{i})}) + n)^{2}}.
\end{aligned}
\label{eq:gamma22}
\end{equation}
Finally, by combining Eqs. (\ref{eq:gamma11}) to (\ref{eq:gamma22}) with Eq. (\ref{eq:CRbound}), one finds the minimum  uncertainties on the parameters:  $\delta \theta_1=\delta a$ and $\delta \theta_2=\delta f_S$.  These are themselves functions of $a$ and $f_S$, so they have to be estimated at a chosen (best-fit) point in the parameter space.

\begin{figure*}[htp]
	\centering
	\includegraphics[width=0.45\textwidth]{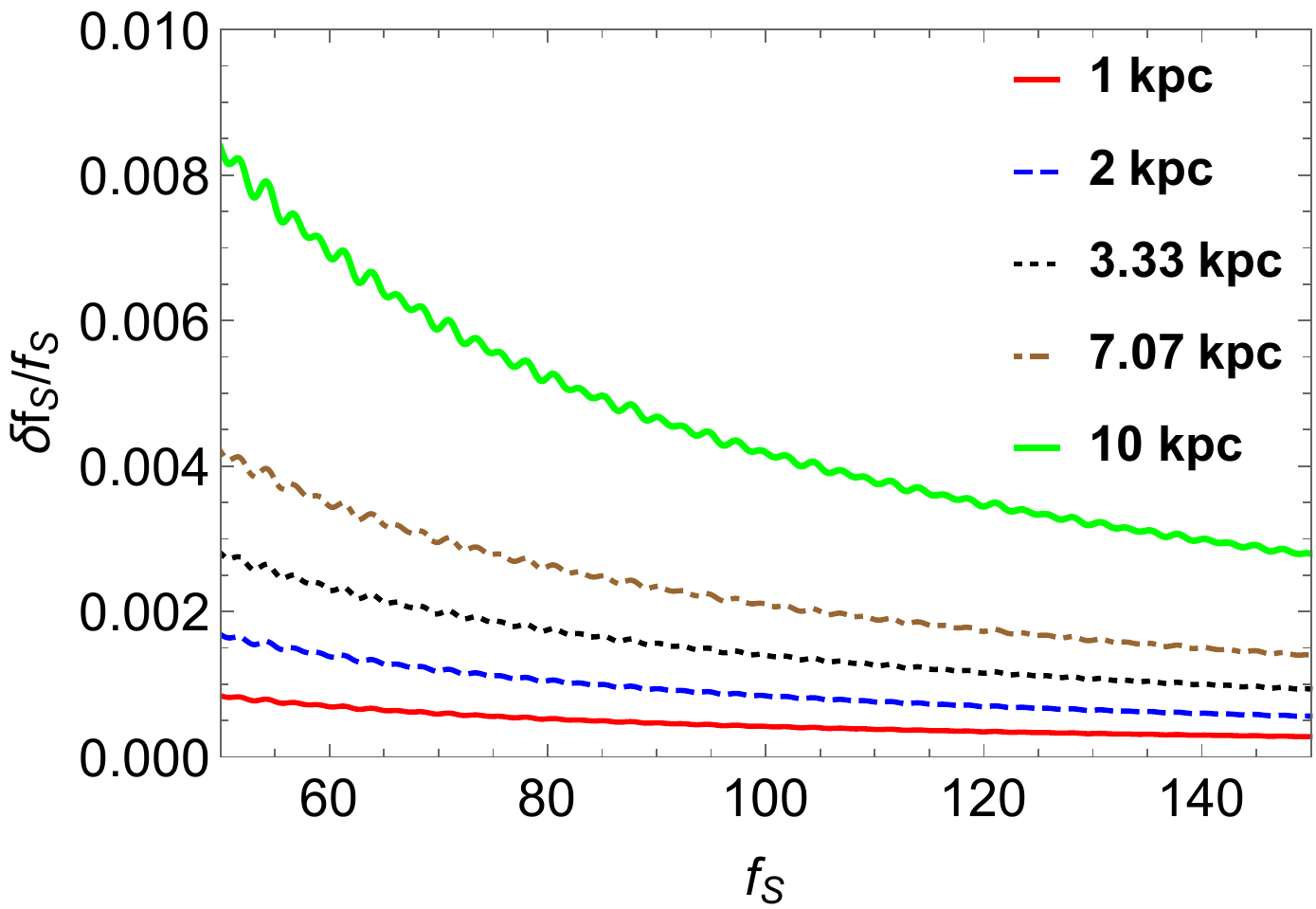}
	\includegraphics[width=0.45\textwidth]{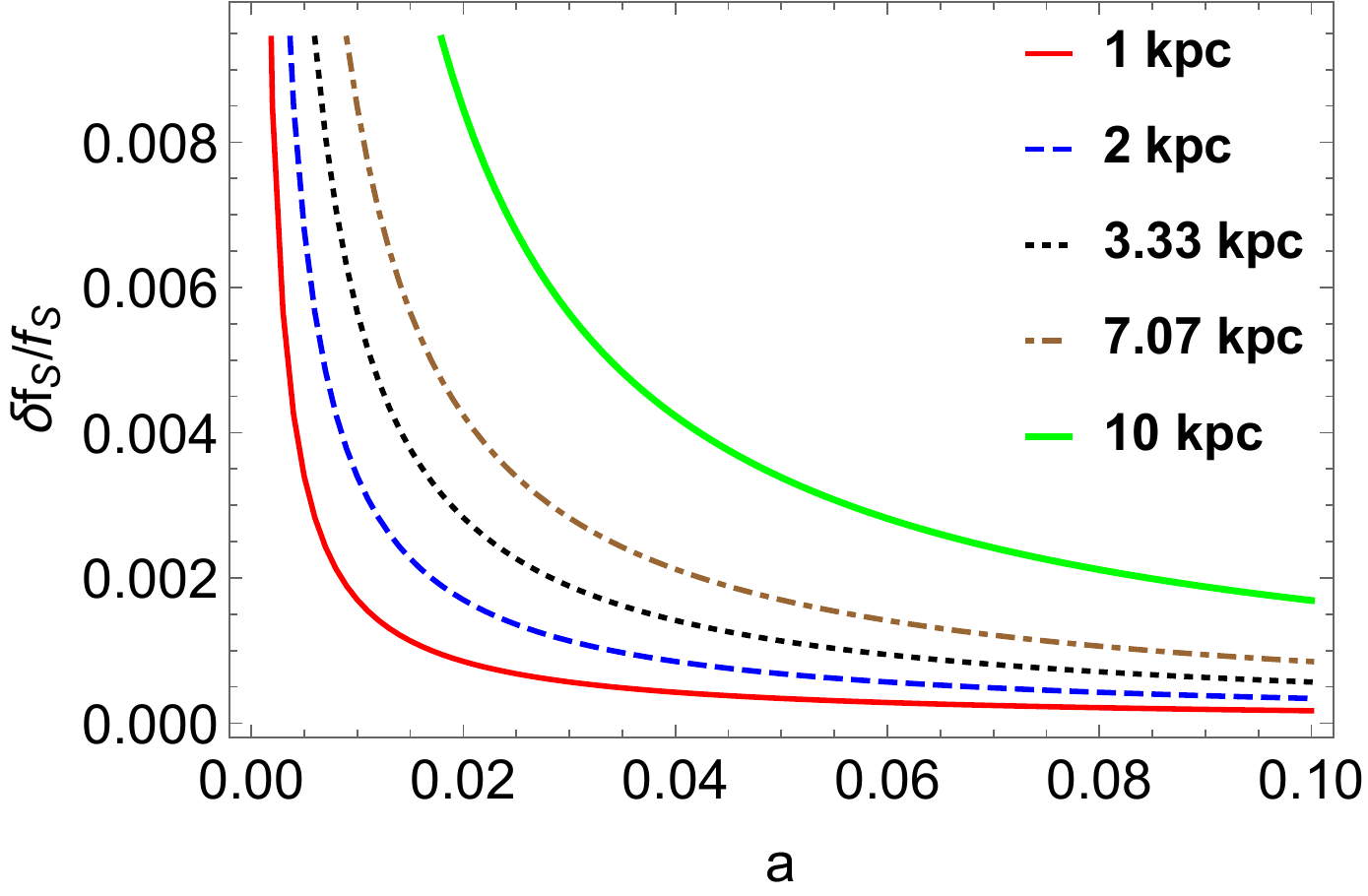} 
	\includegraphics[width=0.45\textwidth]{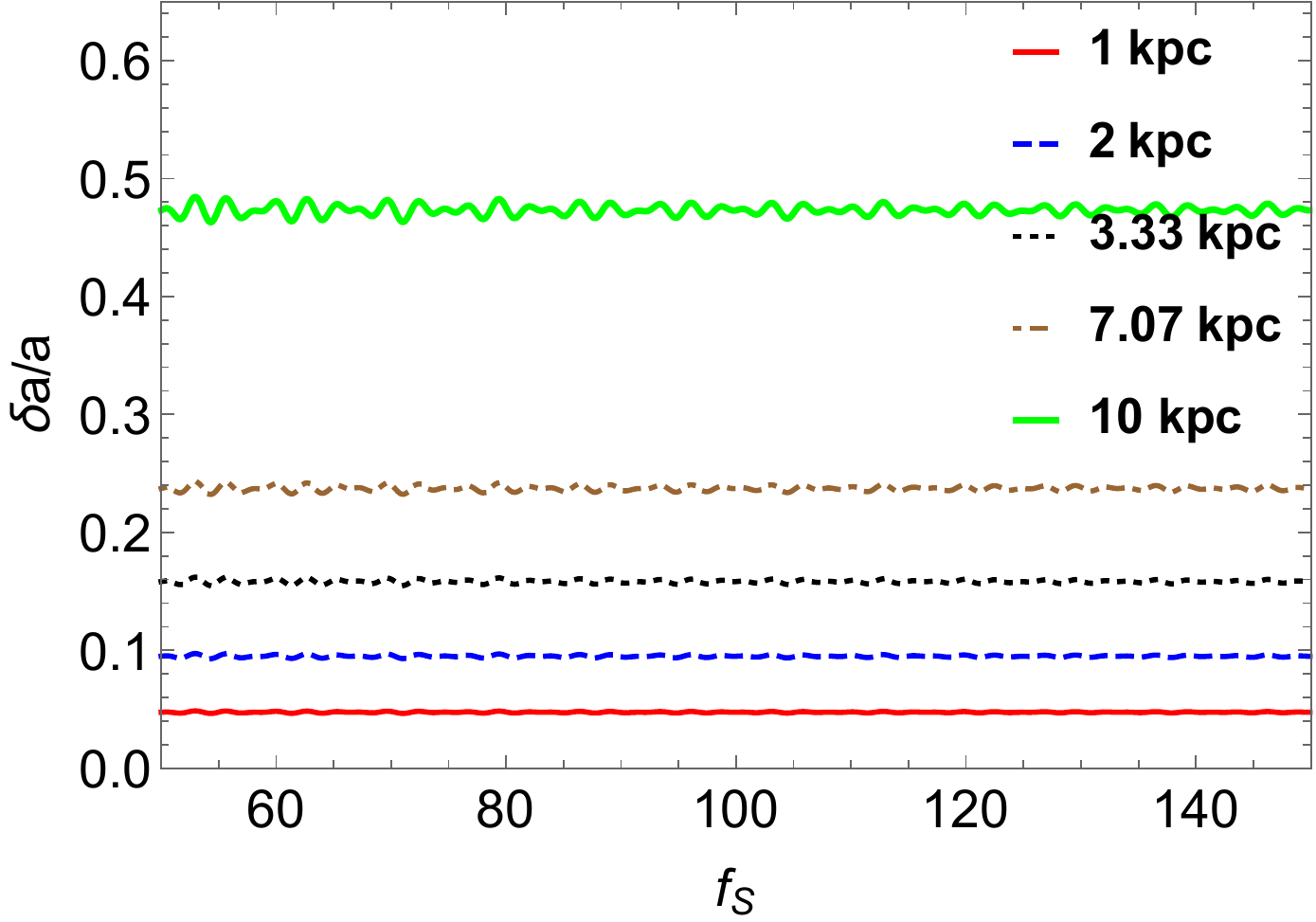}
	\includegraphics[width=0.45\textwidth]{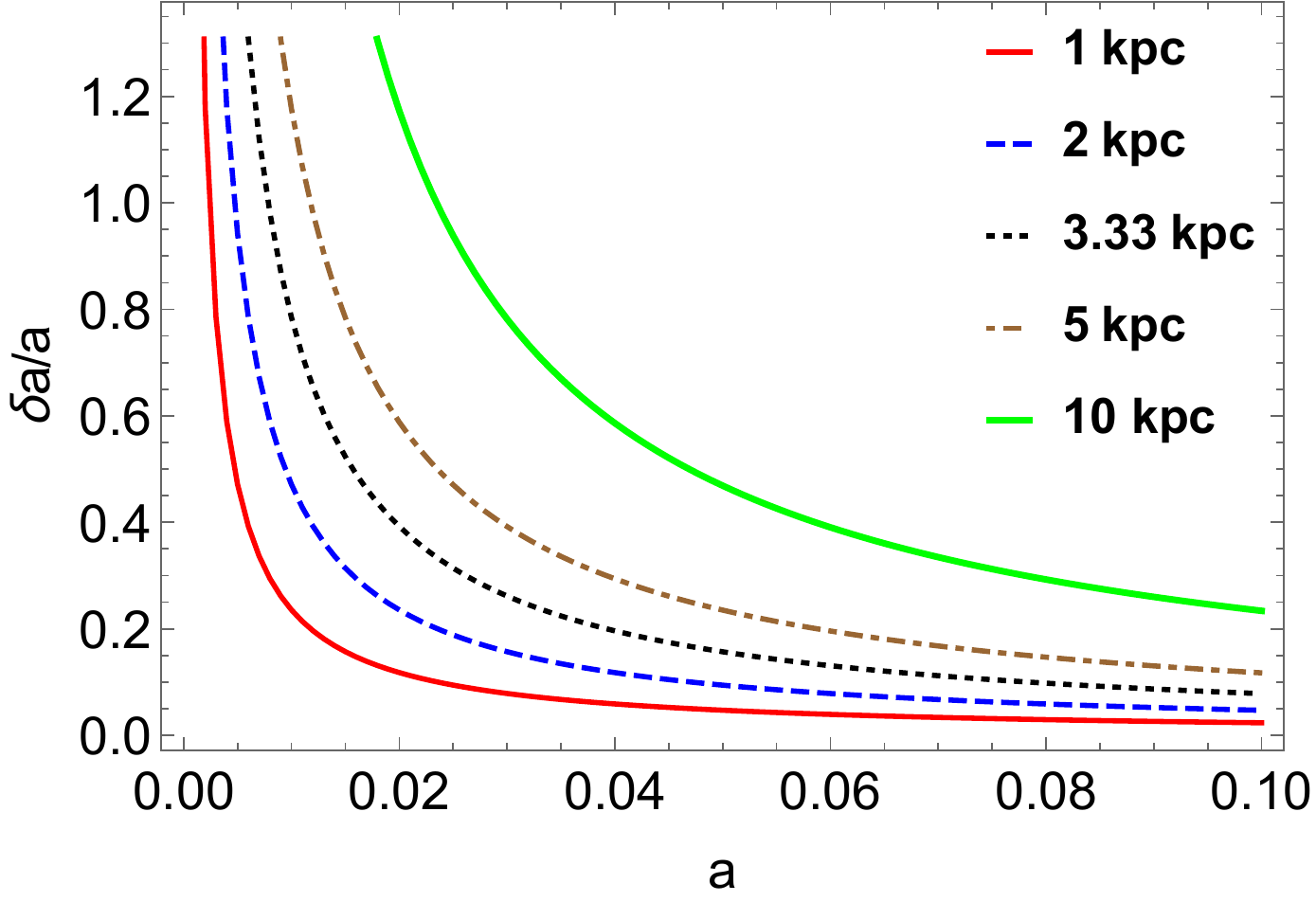} 
    \caption{\cmr\ lower bounds based on simplified parametric templates with SASI (see Eq. \ref{eq:mod2}), in the form of relative errors, for the \si\ frequency (top) and amplitude (bottom), as functions of frequency (left) and amplitude (right),  for \hk\ and select distances to the star (see legend). In each curve, the remaining parameter has been fixed at its best-estimated value (the one for $D=1$ kpc) in Table \ref{tab:sasiPfa10}. } 
	\label{fig:colterFish}
\end{figure*}

\begin{figure*}[htp]
	\centering
	\includegraphics[width=0.45\textwidth]{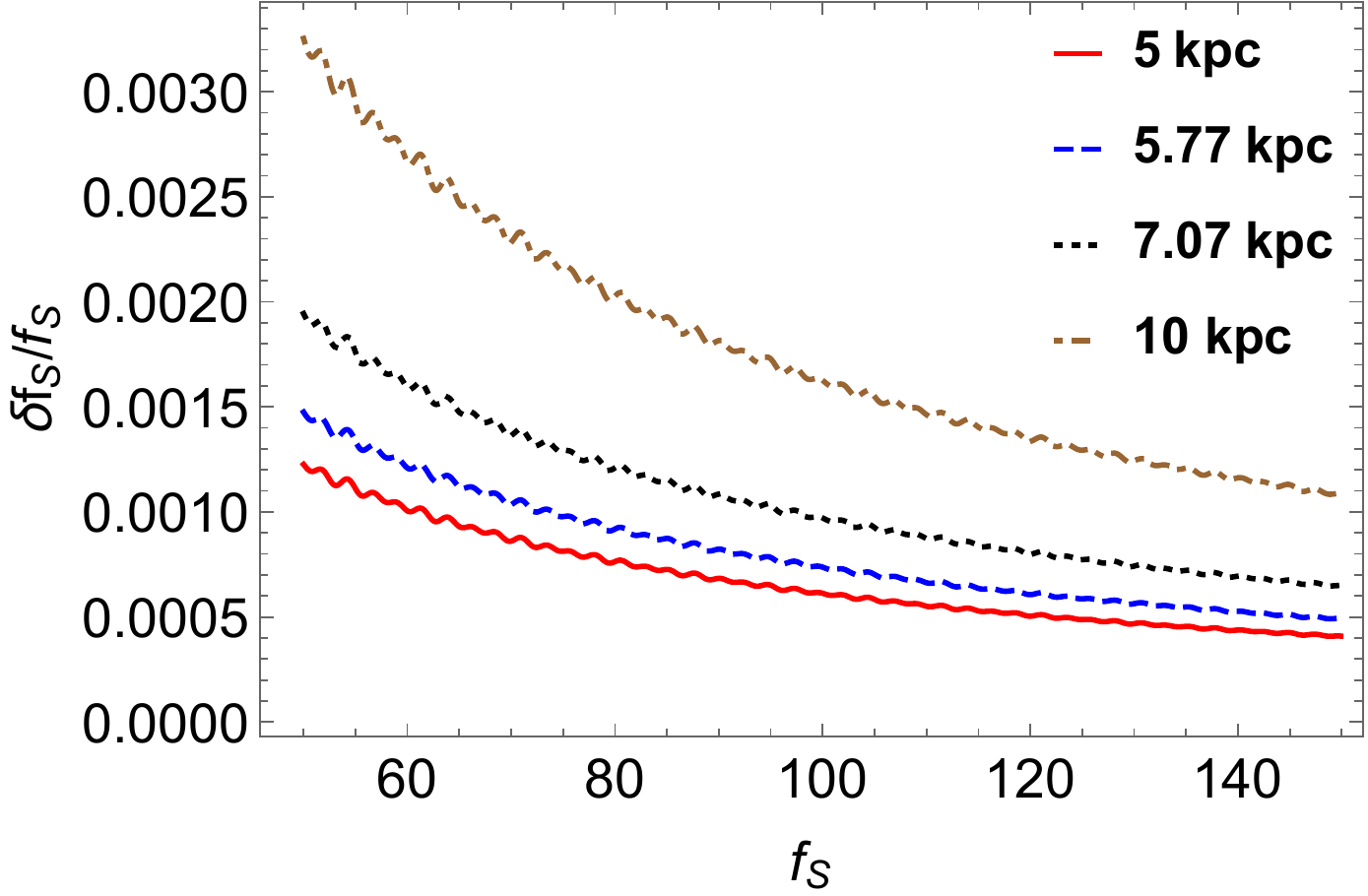}
	\includegraphics[width=0.45\textwidth]{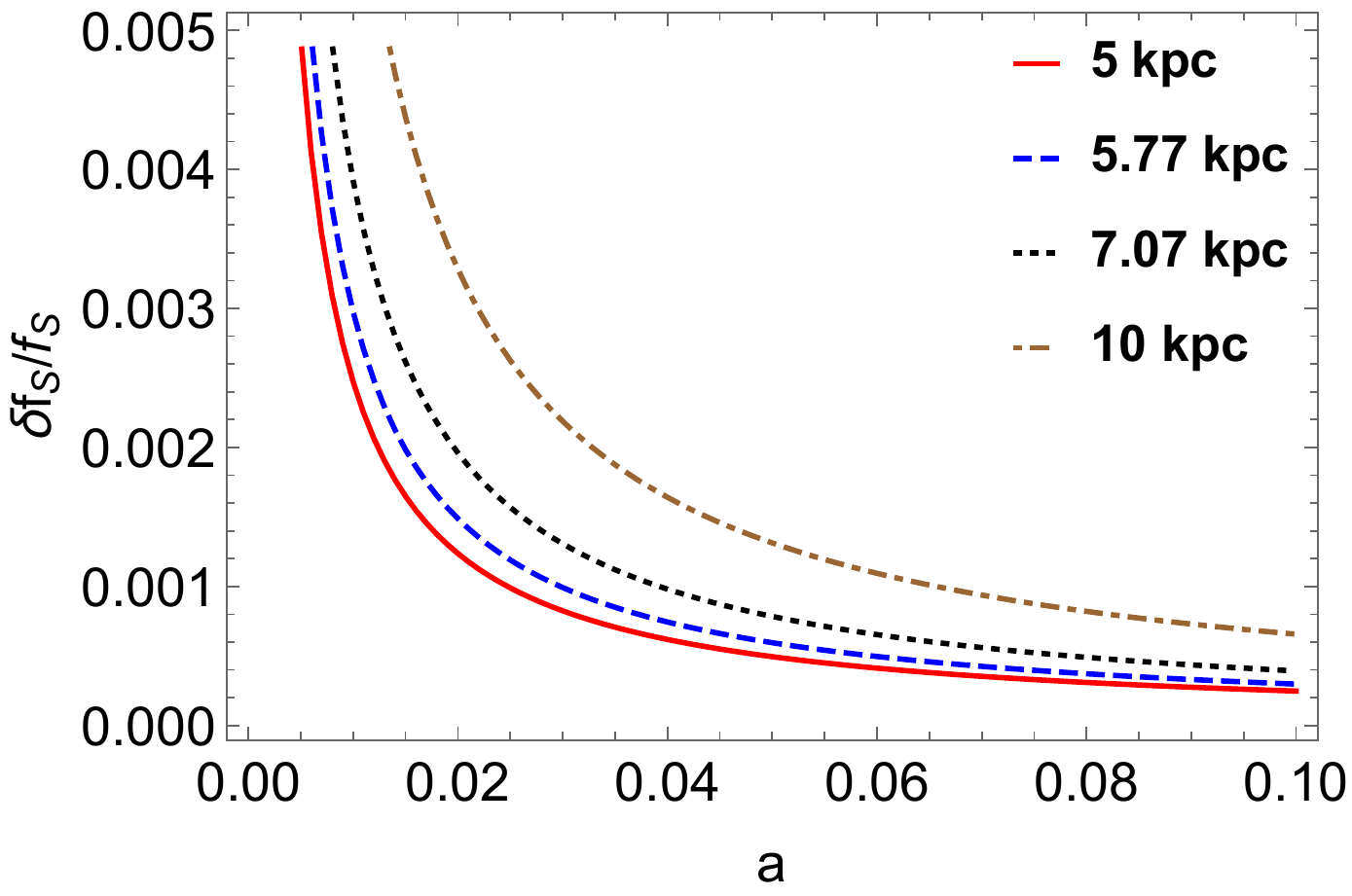} 
	\includegraphics[width=0.45\textwidth]{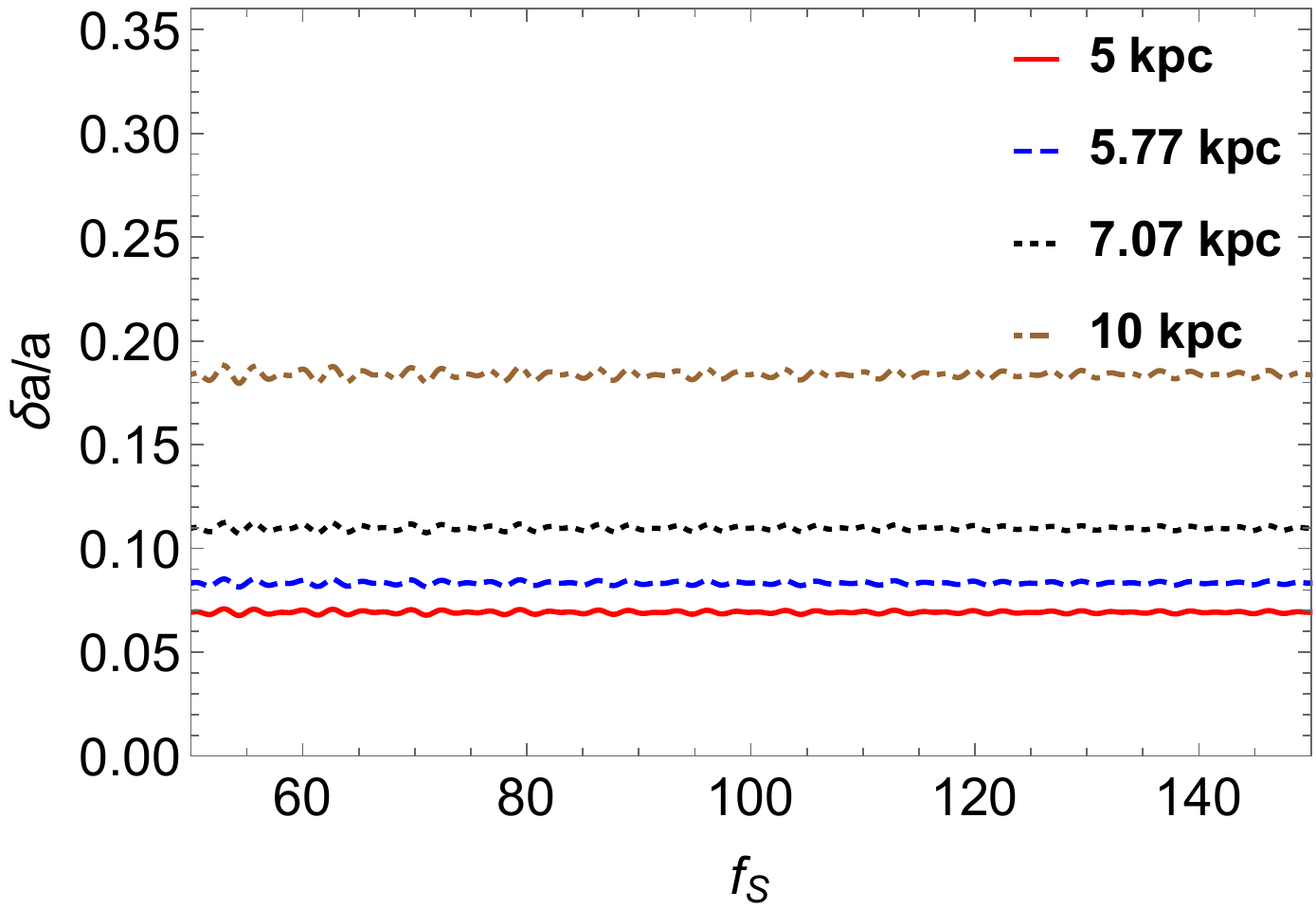}
	\includegraphics[width=0.45\textwidth]{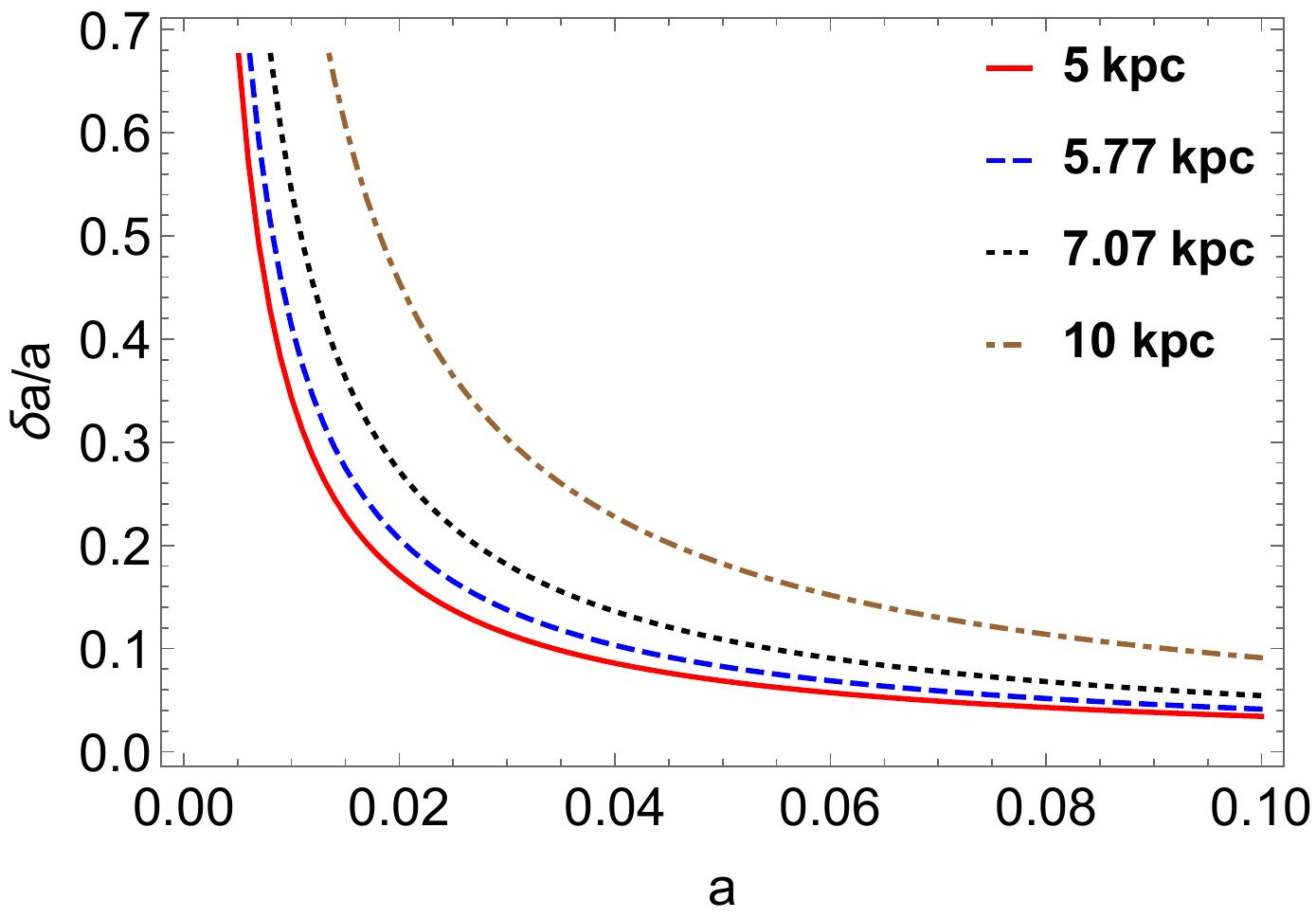} 
	\caption{\cmr\ lower bounds based on simplified parametric templates with SASI (see Eq. \ref{eq:mod2}), in the form of relative errors, for the \si\ frequency (top) and amplitude (bottom), as functions of frequency (left) and amplitude (right),  for \ic\ and select distances to the star (see legend). In each curve, the remaining parameter has been fixed at its best-estimated value (the one for $D=1$ kpc) in Table \ref{tab:sasiIcePfa10}. }
	\label{fig:colterFishIce}
\end{figure*}

In  Figs. \ref{fig:colterFish}-\ref{fig:colterFishIce}, the relative \cmr\ uncertainties are shown for selected distances to the star (which determine the widths of the Gaussian probability distributions that enter the calculation), and as functions of one of the parameters, where the other parameter is kept fixed at its best-estimated value (last columns of Tables \ref{tab:sasiPfa10} and \ref{tab:sasiIcePfa10}). 
As expected, the uncertainties decrease with decreasing distance. We also note that the dependence on the amplitude $a$ is stronger than that on the frequency. 

As a figure of merit, to clarify if our approach is optimal we can
compare the width (error) from the histograms in figure \ref{fig:parameterHypIceC} with
the \cmr\ lower bound for the \si\ analytical model we adopt.
In Tables \ref{tab:sasiPfa10} and \ref{tab:sasiIcePfa10} the \cmr\ uncertainties -- calculated at the points in the parameter space given in the tables themselves --  are listed for the two smallest distances. They can be directly compared to the standard deviations obtained with the likelihood ratio method,  because at such distances the selection effects due to the threshold on ${\mathcal L}$ are negligible (nearly all the simulated cases pass the selection). For larger $D$, where the selection effects on ${\mathcal L}$ are strong, a direct comparison is not meaningful and therefore the \cmr\ bounds are not shown. 

It appears that $\delta a$ obtained from the likelihood ratio is close (sightly larger, as expected) to the corresponding \cmr\ bound, indicating that our method is near optimality for estimating the \si\ amplitude. In contrast, for $\delta f_S$ the \cmr\ bound is orders of magnitude more stringent, so in principle, a more effective method than ours for frequency estimation could exist (although an estimator attaining the Cramer-Rao lower bound does not necessarily exist).   

\section{Summary and discussion}
\label{sec:disc}

We have proposed a novel methodology to do both hypothesis testing and parameter estimation for signatures of \si\ in the time profile of the \n\ event rate from a (galactic) core collapse supernova. 
This method is based on the likelihood ratio constructed using the signal power spectrum, for which the effect of statistical fluctuations was modeled, and suitable frequency cuts can be applied.
%
 We quantify the confidence to identify the presence of SASI in terms of receiver operating curves,
a tool which is commonly used in the gravitational wave community to establish the efficiency versus false alarm probability
for gravitational wave signals  (see, e.g.,  \cite{Blackburn:2005qv}).
 We have tested the effectiveness of the method,  using an injected signal for \hk\ and \ic\ from  a \sn\  numerical simulation by Kuroda,  Kotake,  Hayama and Takiwaki. Specifically, we  have characterized the performance of the method by producing the receiver operating characteristic curves, and by comparing the probability distributions of the best fit parameters (the \si\ frequency and relative amplitude) with the ultimate minimum uncertainties from the \cmr\ lower bounds.

For hypothesis testing, our main result are the probability distributions  in Fig. \ref{fig:likeR}. Figuratively speaking, these can be considered like a calibrated measurement rod against which we will compare the likelihood ratio from an actual, future \sn\ \ns\ detection. We have found that, for a nearby \sn, this ``SASI-meter" is an effective tool: if the experimental likelihood ratio is in the ``red zone" (above a certain threshold for the likelihood ratio, e.g., $\ln {\mathcal L}\gtrsim 30$ for \hk\ and $D\simeq 2$ kpc), then we will be able to confidently claim the presence of \si. If it is in the ``blue zone" ($\ln {\mathcal L}\lesssim 20$ in the same example), then a model without \si\ will be favored, and an upper bound on  the parameters of possible \si\ will be established.  We obtain that, for the KKHT model,  SASI can be identified  with high confidence for a distance to the supernova of up to $\sim 6$ kpc for \ic\ and and up to $\sim 3$ kpc for \hk. The SASI-meter can also be used to identify unusually long periods of SASI, which could help to establish  indication of a failed supernova.

For parameter estimation, we find that, for an injected signal with \si\ and for data sets in the red zone of the SASI-meter, the \si\ frequency and amplitude can be reconstructed if $D \lesssim 5$ kpc for \hk\ ($D \lesssim 10$ kpc for \ic), and their uncertainties are consistent with the \cmr\ lower bounds.
Beyond such distance, the positive response of the SASI-meter, giving indication of an oscillatory pattern in the event rate, is to be attributed to statistical fluctuations and not to the presence of \si.   
The most immediate development of this work will include several three-dimensional supernova simulation results that present SASI and map the performance of the method in different regions of the parameter space. We expect that including several models will result in a blurring of the probability distributions, so the red and blue zones of the SASI-meter will be less clearly separated, or, in other words, the Receiver Operating characteristic Curves will be  worse (i.e., closer to the limiting curve $P_D=P_{FI}$). The  method will remain valid conceptually, however.

In the long term, our goal is to extend the methodology to joint analyses of \n\ and gravitational wave \si\ signals, for a truly multi-messenger approach \cite{IceCube:2018dnn,Branchesi:2016vef,Kalogera:2019bdd}. Within this goal, the present paper serves to create the foundation of a formalism for \ns\ that finds a direct counterpart (using the same tools, like the likelihood ratio and the receiver operating characteristic curve, for example) in existing gravitational wave analysis protocols. Moving forward, new approaches will have to be developed to establish how to most effectively  combine the two signals, \ns\ and gravitational waves, that have both similarities (e.g., similar \si\ frequency) and important differences (different sources of noise, for example). Such development work will explore further the territory of multi-messenger astronomy and aid the investigation of a future galactic core collapse supernova.

\subsubsection*{Acknowledgments}
We are thankful to Takami Kuroda who kindly provided the \n\ event rates required for the 
analysis of this work.
CL and ZL acknowledge funding from the National Science Foundation grant number PHY-1613708.
KK thanks Tomoya Takiwaki for stimulating discussions and acknowledges support by Grant-in-Aid for Scientific Research 
(JP17H01130)
from the Japan Society for Promotion  of Science (JSPS) and the Ministry of Education, Science and Culture of Japan (MEXT, Nos. JP17H06357, 
JP17H06364), and by the Central Research Institute of Stellar Explosive Phenomena (REISEP) at Fukuoka University and the associated projects (Nos.\ 171042,177103).
MZ acknowledges funding from the National Science Foundation grant number PHY-1806885.
\appendix
\section{Probability distribution of Fourier transformed neutrino signal}
In this appendix the probability distribution of a Fourier transformed neutrino signal in frequency domain defined in eq. (\ref{eq:h}) is given.
We start from the real part of $\tilde{h}$: 
\begin{equation}
	Re(\tilde{h})=\sum_{j=0}^{N_{bins}-1}\tilde{N}(t_j)\cos(2\pi t_j k\delta),
\label{eq:reh}
\end{equation}
where $\tilde{N}$ is the observed event number in a time bin, statistically fluctuating around its mean value $N$.

To simplify the notation, let us define $\tilde{N}_j \equiv \tilde{N}(t_j)$, $\tilde{n}_j \equiv \tilde{N_j}\cos(2\pi t_j k\delta) $, and  $\tilde{h}_R \equiv Re(\tilde{h})$. 
Here $n_j$ and $N_j$ will be means of $\tilde{n}_j$ and $\tilde{N_j}$ respectively,  so that  ${n}_j = {N_j}\cos(2\pi t_j k\delta) $.
Since the neutrino event number $\tilde{N_j}=\tilde{R}(t_j)\Delta$ follows a Gaussian distribution (with variance $\sqrt{N_j}$),  $\tilde{n}_j$ also follows a Gaussian distribution, with variance $\sqrt{N_j}\cos(2\pi t_j k\delta)$. Specifically, the two distributions are:
\begin{equation}
	Prob(\tilde{N}_j)=\frac{1}{\sqrt{2\pi N_j}}e^{-\frac{(\tilde{N}_j-N_j)^2}{2N_j}},
\end{equation}
\begin{eqnarray}
	Prob(\tilde{n}_j)
	=\frac{1}{\sqrt{2\pi N_j \cos^2(2\pi t_j k\delta )}}e^{-\frac{(\tilde{n}_j-n_j)^2}{2 N_j \cos^2(2\pi t_j k\delta )}}~. 
\end{eqnarray}
Below we show  that the Fourier transformed neutrino signal in frequency domain, which is a sum of Gaussian distributed random values, follows a Gaussian distribution as well. The probability distribution for $\tilde{h}_R$ is defined as:
\begin{equation}
	Prob(\tilde{h}_R)=\int(\prod_{j=0}^{N_{bins}-1}Prob(\tilde{n}_j))\delta(\tilde{h}_R-\sum_{j=0}^{N_{bins}-1}\tilde{n}_j)d\tilde{n}_0...d\tilde{n}_j.
\end{equation} 
Let us now perform a Fourier transform of eq. (A4):
\begin{equation}
\begin{split}
\int Prob(\tilde{h}_R)e^{il\tilde{h}_R}d\tilde{h}_R=&\int(\prod_{j=0}^{N_{bins}-1}Prob(\tilde{n}_j))\\&\times e^{il\sum_{j=0}^{N_{bins}-1}\tilde{n}_j}d\tilde{n}_0...d\tilde{n}_j\\=&\prod_{j=0}^{N_{bins}-1}\int Prob(\tilde{n}_j)e^{il\tilde{n}_j}d\tilde{n}_j~,
\end{split}
\end{equation}
and note that the Fourier transform of a Gaussian distribution $P_G(x)$ with mean value $\mu$ and standard deviation $\sigma$ is:
\begin{equation}
\int P_G(x)e^{ilx}dx=e^{il\mu}e^{-\frac{\sigma^2l^2}{2}}~.
\end{equation}
Then, we get:
\begin{equation}
\begin{split}
	\int Prob(\tilde{h}_R)e^{il\tilde{h}_R}d\tilde{h}_R=&e^{il\sum_{j=0}^{N_{bins}-1}n_j}\\&\times e^{-l^2\frac{\sum_{j=0}^{N_{bins}-1}n^2_j}{2N_j}}.
\end{split}
\end{equation}
We then do an inverse Fourier transformation of eq. (A7), and obtain:
\begin{equation}
\begin{split}
Prob(\tilde{h}_R)=&\frac{1}{\sqrt{2\pi\sum_{j=0}^{N_{bins}-1}N_j \cos^2(2\pi t_j k\delta )}}\\&\times e^{\frac{-(\tilde{h}_R-\sum_{j=0}^{N_{bins}-1}n_j)^2}{2\sum_{j=0}^{N_{bins}-1}N_j \cos^2(2\pi t_j k\delta )}}\\=&\frac{1}{\sqrt{2\pi\sigma_R^2}}e^{\frac{(\tilde{h}_R-h_R)}{2\sigma_R^2}},
\end{split}
\label{eq:imgproof}
\end{equation}
with $\sigma_R^2=\sum_{j=0}^{N_{bins}-1}N_j \cos^2(2\pi t_j k\delta )$ and $h_R=\sum_{j=0}^{N_{bins}-1}n_j$.

Eq. (\ref{eq:imgproof}) concludes the proof for the real part of $\tilde h$. A similar proof can be done for the imaginary part of $\tilde h$, with the replacement $\cos(2\pi t_j k\delta) \rightarrow \sin(2\pi t_j k\delta)$, leading to a result analogous to Eq. (\ref{eq:imgproof}).

Let us now prove the statistical independence of the values of $\tilde h$ in different freuquency bins.  
First, it is known that neutrino event rates in each time bin are statistically independent, i.e.:

\begin{equation}
\langle \tilde{N}(t_1)\tilde{N}(t_2)\rangle=\langle \tilde{N}(t_1)\rangle\langle \tilde{N}(t_2)\rangle,
\end{equation}
where $t_1\neq t_2$.
It then follows that:
\begin{equation}
\label{indprof}
\begin{split}
	\langle \tilde{h}(k\delta)\tilde{h}^*(k'\delta)\rangle&=\sum_{m}^{N_{bins}}\sum_{l}^{N_{bins}}e^{it_l k\delta}\langle \tilde{N}(t_l)\tilde{N}(t_m)\rangle e^{-it_m k'\delta}\\&=\sum_{l}^{N_{bins}}e^{it_l k\delta}\langle \tilde{N}(t_l)\rangle\sum_{m}^{N_{bins}}\langle \tilde{N}(t_m)\rangle e^{-it_m k'\delta}\\&+\sum_{l}^{N_{bins}}N(t_l) e^{-it_l(k-k')\delta}\\&=\langle \tilde{h}(k\delta)\rangle \langle \tilde{h}^*(k'\delta)\rangle+\sum_{l}^{N_{bins}}N(t_l) e^{-it_l(k-k')\delta}~. 
\end{split}
\end{equation}
The second term on the right hand side of Eq. (\ref{indprof}) comes from the contribution of the terms  with $l=m$, and is much smaller than the first term given that $N\gg 10$ and $N_{bins}\gg 10$.
Therefore, Eq. (\ref{indprof}) shows an approximate statistical independence for $\tilde h$ in different frequency bins. 
The same conclusion can be reached for the real and imaginary parts of $\tilde h$ ($\tilde h=h_R+ i h_I$) by rewriting Eq. (\ref{indprof}) in terms of $h_R$ and $h_I$.

\section{Probability distribution of Power}

In this appendix we derive Eq. (\ref{eq:prob}). The power at a specific frequency is written as:
\begin{equation}
\label{eq:power}
	\tilde{P}=C(\tilde{h}_R^2+\tilde{h}_I^2),
\end{equation} 
where $C=2/N_{bins}^2$ is the normalization factor from Eq. (\ref{eq:power1}).
It can be shown that the standard deviations of $\tilde{h}_R$ and $\tilde{h}_I$, denoted as  $\sigma_R^2$ and $\sigma_I^2$ respectively, are approximately equal, with  
differences of less than 10\% (we checked this by comparing the two quantities  in the whole  space of the parameters $\Omega=\{a,f_S\}$). Accordingly in the following we assume that:
\begin{equation}
\begin{split}
	\sigma_R^2\approxeq\sigma_I^2\approxeq\sigma^2&\equiv\frac{\sigma_I^2+\sigma_R^2}{2}\\&=\sum_{j=0}^{N_{bins}-1}\frac{N_j(\cos^2(2\pi t_j k\delta)+sin^2(2\pi t_j k\delta ))}{2}\\&=\frac{N_{ev}}{2}~. 
	\label{eq:sigmaaprox}
\end{split}
\end{equation}
We now define new variable $\tilde{P}'$ :
\begin{equation}
	\tilde{P}'=\frac{\tilde{P}}{C\sigma^2}=\frac{\tilde{h}_R^2+\tilde{h}_I^2}{\sigma^2}.
\end{equation}
Note that both $\tilde{h}_R/\sigma$ and $\tilde{h}_I/\sigma$ are Gaussian random variables with unit standard deviation given Eq. (\ref{eq:sigmaaprox}). As a result,
$\tilde{P}'$ follows a non-central-chi-squared distribution \cite{2009fundamentals}:
\begin{equation}
	Prob(\tilde{P}')=\frac{1}{2}e^{\frac{1}{2}(\tilde{P}'+\lambda')}I_0 \left(\sqrt{\lambda'\tilde{P}'} \right),
\end{equation}
where $\lambda'=(h_R^2+h_I^2)/\sigma^2$ is the noncentrality parameter and $I_0$ is the modified Bessel function of the first kind.
By using the normalization condition:
\begin{equation}
	\int Prob(\tilde{P}')d\tilde{P}'=\int Prob(\tilde{P})d\tilde{P}=1,
\end{equation}
the probability density function of $\tilde{P}$ is:
\begin{equation}
\label{eq:powerProb}
	Prob(\tilde{P})=\frac{1}{2}e^{-\frac{1}{2}\left(\frac{\tilde{P}}{C\sigma^2}+\lambda'\right)}I_0 \left(\sqrt{\lambda'\frac{\tilde{P}}{C\sigma^2}} \right)\frac{1}{C\sigma^2}.
\end{equation}
We then insert eq. (\ref{eq:power}) and eq. (\ref{eq:sigmaaprox}) into eq. (\ref{eq:powerProb}) to get the probability density function as written in eq. (\ref{eq:prob}). The analytical expression for the probability density distribution of power agrees very well with our numerical Monte Carlo simulation. 

\section{Fisher Matrix}

In this appendix, we show that Eq. (\ref{eq:gamma2}) follows from Eq. (\ref{eq:gamma}).

Integrating by parts with null boundary condition, eq. (\ref{eq:gamma}) can be re-written as:
\begin{equation}
\begin{split}
	\Gamma_{\alpha\beta}&=\int-\frac{\partial^2 \ln{Prob(\tilde{R})}}{\partial\theta_{\alpha}\theta_{\beta}}Prob(\tilde{R})d\tilde{R}\\&=\int\frac{\partial \ln{Prob(\tilde{R})}}{\partial\theta_{\alpha}}\frac{\partial \ln{Prob(\tilde{R})}}{\partial\theta_{\beta}}Prob(\tilde{R})d\tilde{R}~. 
\end{split}
\end{equation}
Using the expression: 
\begin{equation}
\ln{Prob(\tilde{R})}=\sum_i \left(-\frac{1}{2}\ln(2\pi\sigma_i^2)-\frac{(\tilde{R}_i-R_i)^2}{2\sigma_i^2}\right)~, 
\end{equation}
we then obtain:
\begin{equation}
\begin{split}
	\frac{\partial \ln{Prob(\tilde{R})}}{\partial\theta_{\alpha}}&=\sum_i\frac{1}{\sigma^2_i}(\tilde{R}_i-R_i)\frac{\partial R_i}{\partial\theta_{\alpha}}\\&+\sum_i \left(-\frac{1}{2\sigma^2_i}\frac{\partial\sigma_i^2}{\partial\theta_{\alpha}}+\frac{(\tilde R_i-R_i)^2}{2\sigma^4_i}\left(\frac{\partial\sigma^2_i}{\partial\theta_{\alpha}}\right) \right).
\end{split}
\end{equation}
The integration in Eq. (C1) is then divided into several parts, and becomes:

\begin{widetext}

\begin{equation}
\begin{split}
\int\frac{\partial\ln{Prob(\tilde{R})}}{\partial\theta_{\alpha}}\frac{\partial\ln{Prob(\tilde{R})}}{\partial\theta_{\beta}}Prob(\tilde{R})d\tilde{R}&=
\sum_{i}\frac{\partial R_i}{\partial\theta_{\alpha}}\frac{\partial R_j}{\partial\theta_{\beta}}\frac{1}{\sigma_i^2}
+\sum_{i,j}\left(\frac{1}{2\sigma_i^2}\frac{\partial\sigma_i^2}{\partial\theta_{\alpha}}\right)\left(\frac{1}{2\sigma_j^2}\frac{\partial\sigma_j^2}{\partial\theta_{\beta}}\right)\\
&+\int d\tilde{R}\sum_{i,j}\frac{(\tilde{R}_i-R_i)^2}{2\sigma_i^4}\frac{\partial\sigma^2_i}{\partial\theta_{\alpha}}\frac{(\tilde{R}_j-R_j)^2}{2\sigma_j^4}\frac{\partial\sigma^2_j}{\partial\theta_{\beta}}Prob(\tilde{R})\\
&+\int d\tilde{R}\sum_{i,j}\left(-\frac{1}{2\sigma_i^2}\frac{\partial\sigma^2_i}{\partial\theta_{\alpha}}\right)\frac{(\tilde{R}_j-R_j)^2}{2\sigma_j^4}\frac{\partial\sigma^2_j}{\partial\theta_{\beta}}Prob(\tilde{R})
+\{i\leftrightarrow j\}~.
\end{split}
\label{eq:c7}
\end{equation}
\end{widetext}

Note that:
\begin{equation}
	\langle(\tilde{R}-R)^4\rangle=3\sigma^4,
\end{equation}
and that the second and the fourth term in Eq. (\ref{eq:c7}) sum to zero. Finally eq. (\ref{eq:c7}) becomes:
\begin{equation}
\begin{split}
\Gamma_{\alpha\beta}&=\sum_{i}\frac{\partial R_i}{\partial\theta_{\alpha}}\frac{\partial R_j}{\partial\theta_{\beta}}\frac{1}{\sigma_i^2}+\frac{1}{2}\sum_i\frac{1}{\sigma_i^2}\frac{\partial\sigma^2_i}{\partial\theta_{\alpha}}\frac{1}{\sigma_i^2}\frac{\partial\sigma^2_i}{\partial\theta_{\theta}}\\&=\mu_{\alpha}^{T} \Sigma^{-1} \mu_{\beta} + \frac{1}{2} Tr[\Tilde{c}_{\alpha} \Tilde{c}_{\beta}]~,
\end{split}
\end{equation}
 which concludes the proof. 


\bibliography{b}

\begin{thebibliography}{51}
\expandafter\ifx\csname natexlab\endcsname\relax\def\natexlab#1{#1}\fi
\expandafter\ifx\csname bibnamefont\endcsname\relax
  \def\bibnamefont#1{#1}\fi
\expandafter\ifx\csname bibfnamefont\endcsname\relax
  \def\bibfnamefont#1{#1}\fi
\expandafter\ifx\csname citenamefont\endcsname\relax
  \def\citenamefont#1{#1}\fi
\expandafter\ifx\csname url\endcsname\relax
  \def\url#1{\texttt{#1}}\fi
\expandafter\ifx\csname urlprefix\endcsname\relax\def\urlprefix{URL }\fi
\providecommand{\bibinfo}[2]{#2}
\providecommand{\eprint}[2][]{\url{#2}}

\bibitem[{\citenamefont{Blondin et~al.}(2003)\citenamefont{Blondin, Mezzacappa,
  and DeMarino}}]{Blondin:2002sm}
\bibinfo{author}{\bibfnamefont{J.~M.} \bibnamefont{Blondin}},
  \bibinfo{author}{\bibfnamefont{A.}~\bibnamefont{Mezzacappa}},
  \bibnamefont{and} \bibinfo{author}{\bibfnamefont{C.}~\bibnamefont{DeMarino}},
  \bibinfo{journal}{Astrophys. J.} \textbf{\bibinfo{volume}{584}},
  \bibinfo{pages}{971} (\bibinfo{year}{2003}).

\bibitem[{\citenamefont{Foglizzo et~al.}(2015)}]{Foglizzo:2015dma}
\bibinfo{author}{\bibfnamefont{T.}~\bibnamefont{Foglizzo}}
  \bibnamefont{et~al.}, \bibinfo{journal}{Publ. Astron. Soc. Austral.}
  \textbf{\bibinfo{volume}{32}}, \bibinfo{pages}{e009} (\bibinfo{year}{2015}).

\bibitem[{\citenamefont{Mirizzi et~al.}(2016)\citenamefont{Mirizzi, Tamborra,
  Janka, Saviano, Scholberg, Bollig, Hudepohl, and
  Chakraborty}}]{Mirizzi:2015eza}
\bibinfo{author}{\bibfnamefont{A.}~\bibnamefont{Mirizzi}},
  \bibinfo{author}{\bibfnamefont{I.}~\bibnamefont{Tamborra}},
  \bibinfo{author}{\bibfnamefont{H.-T.} \bibnamefont{Janka}},
  \bibinfo{author}{\bibfnamefont{N.}~\bibnamefont{Saviano}},
  \bibinfo{author}{\bibfnamefont{K.}~\bibnamefont{Scholberg}},
  \bibinfo{author}{\bibfnamefont{R.}~\bibnamefont{Bollig}},
  \bibinfo{author}{\bibfnamefont{L.}~\bibnamefont{Hudepohl}}, \bibnamefont{and}
  \bibinfo{author}{\bibfnamefont{S.}~\bibnamefont{Chakraborty}},
  \bibinfo{journal}{Riv. Nuovo Cim.} \textbf{\bibinfo{volume}{39}},
  \bibinfo{pages}{1} (\bibinfo{year}{2016}).

\bibitem[{\citenamefont{Kuroda et~al.}(2017)\citenamefont{Kuroda, Kotake,
  Hayama, and Takiwaki}}]{Kuroda:2017trn}
\bibinfo{author}{\bibfnamefont{T.}~\bibnamefont{Kuroda}},
  \bibinfo{author}{\bibfnamefont{K.}~\bibnamefont{Kotake}},
  \bibinfo{author}{\bibfnamefont{K.}~\bibnamefont{Hayama}}, \bibnamefont{and}
  \bibinfo{author}{\bibfnamefont{T.}~\bibnamefont{Takiwaki}},
  \bibinfo{journal}{Astrophys. J.} \textbf{\bibinfo{volume}{851}},
  \bibinfo{pages}{62} (\bibinfo{year}{2017}).

\bibitem[{\citenamefont{Andresen et~al.}(2017)\citenamefont{Andresen, Müller,
  Müller, and Janka}}]{Andresen:2016pdt}
\bibinfo{author}{\bibfnamefont{H.}~\bibnamefont{Andresen}},
  \bibinfo{author}{\bibfnamefont{B.}~\bibnamefont{Müller}},
  \bibinfo{author}{\bibfnamefont{E.}~\bibnamefont{Müller}}, \bibnamefont{and}
  \bibinfo{author}{\bibfnamefont{H.-T.} \bibnamefont{Janka}},
  \bibinfo{journal}{Mon. Not. Roy. Astron. Soc.}
  \textbf{\bibinfo{volume}{468}}, \bibinfo{pages}{2032} (\bibinfo{year}{2017}).

\bibitem[{\citenamefont{{M{\"u}ller} and {Janka}}(2014)}]{bernhard14}
\bibinfo{author}{\bibfnamefont{B.}~\bibnamefont{{M{\"u}ller}}}
  \bibnamefont{and} \bibinfo{author}{\bibfnamefont{H.-T.}
  \bibnamefont{{Janka}}}, \bibinfo{journal}{\apj}
  \textbf{\bibinfo{volume}{788}}, \bibinfo{eid}{82} (\bibinfo{year}{2014}).

\bibitem[{\citenamefont{Blondin and Shaw}(2007)}]{Blondin:2006fx}
\bibinfo{author}{\bibfnamefont{J.~M.} \bibnamefont{Blondin}} \bibnamefont{and}
  \bibinfo{author}{\bibfnamefont{S.}~\bibnamefont{Shaw}},
  \bibinfo{journal}{Astrophys. J.} \textbf{\bibinfo{volume}{656}},
  \bibinfo{pages}{366} (\bibinfo{year}{2007}).

\bibitem[{\citenamefont{Marek et~al.}(2009)\citenamefont{Marek, Janka, and
  Mueller}}]{Marek:2008qi}
\bibinfo{author}{\bibfnamefont{A.}~\bibnamefont{Marek}},
  \bibinfo{author}{\bibfnamefont{H.~T.} \bibnamefont{Janka}}, \bibnamefont{and}
  \bibinfo{author}{\bibfnamefont{E.}~\bibnamefont{Mueller}},
  \bibinfo{journal}{Astron. Astrophys.} \textbf{\bibinfo{volume}{496}},
  \bibinfo{pages}{475} (\bibinfo{year}{2009}).

\bibitem[{\citenamefont{Marek and Janka}(2009)}]{Marek:2007gr}
\bibinfo{author}{\bibfnamefont{A.}~\bibnamefont{Marek}} \bibnamefont{and}
  \bibinfo{author}{\bibfnamefont{H.~T.} \bibnamefont{Janka}},
  \bibinfo{journal}{Astrophys. J.} \textbf{\bibinfo{volume}{694}},
  \bibinfo{pages}{664} (\bibinfo{year}{2009}).

\bibitem[{\citenamefont{Nakamura et~al.}(2015)\citenamefont{Nakamura, Takiwaki,
  Kuroda, and Kotake}}]{Nakamura:2014caa}
\bibinfo{author}{\bibfnamefont{K.}~\bibnamefont{Nakamura}},
  \bibinfo{author}{\bibfnamefont{T.}~\bibnamefont{Takiwaki}},
  \bibinfo{author}{\bibfnamefont{T.}~\bibnamefont{Kuroda}}, \bibnamefont{and}
  \bibinfo{author}{\bibfnamefont{K.}~\bibnamefont{Kotake}},
  \bibinfo{journal}{Publ. Astron. Soc. Jap.} \textbf{\bibinfo{volume}{67}},
  \bibinfo{pages}{107} (\bibinfo{year}{2015}).

\bibitem[{\citenamefont{Summa et~al.}(2016)\citenamefont{Summa, Hanke, Janka,
  Melson, Marek, and Müller}}]{Summa:2015nyk}
\bibinfo{author}{\bibfnamefont{A.}~\bibnamefont{Summa}},
  \bibinfo{author}{\bibfnamefont{F.}~\bibnamefont{Hanke}},
  \bibinfo{author}{\bibfnamefont{H.-T.} \bibnamefont{Janka}},
  \bibinfo{author}{\bibfnamefont{T.}~\bibnamefont{Melson}},
  \bibinfo{author}{\bibfnamefont{A.}~\bibnamefont{Marek}}, \bibnamefont{and}
  \bibinfo{author}{\bibfnamefont{B.}~\bibnamefont{Müller}},
  \bibinfo{journal}{Astrophys. J.} \textbf{\bibinfo{volume}{825}},
  \bibinfo{pages}{6} (\bibinfo{year}{2016}).

\bibitem[{\citenamefont{Blondin and Mezzacappa}(2007)}]{Blondin:2006yw}
\bibinfo{author}{\bibfnamefont{J.~M.} \bibnamefont{Blondin}} \bibnamefont{and}
  \bibinfo{author}{\bibfnamefont{A.}~\bibnamefont{Mezzacappa}},
  \bibinfo{journal}{Nature} \textbf{\bibinfo{volume}{445}}, \bibinfo{pages}{58}
  (\bibinfo{year}{2007}).

\bibitem[{\citenamefont{Iwakami et~al.}(2009)\citenamefont{Iwakami, Kotake,
  Ohnishi, Yamada, and Sawada}}]{Iwakami:2008qj}
\bibinfo{author}{\bibfnamefont{W.}~\bibnamefont{Iwakami}},
  \bibinfo{author}{\bibfnamefont{K.}~\bibnamefont{Kotake}},
  \bibinfo{author}{\bibfnamefont{N.}~\bibnamefont{Ohnishi}},
  \bibinfo{author}{\bibfnamefont{S.}~\bibnamefont{Yamada}}, \bibnamefont{and}
  \bibinfo{author}{\bibfnamefont{K.}~\bibnamefont{Sawada}},
  \bibinfo{journal}{Astrophys. J.} \textbf{\bibinfo{volume}{700}},
  \bibinfo{pages}{232} (\bibinfo{year}{2009}).

\bibitem[{\citenamefont{Fernandez}(2010)}]{Fernandez:2010db}
\bibinfo{author}{\bibfnamefont{R.}~\bibnamefont{Fernandez}},
  \bibinfo{journal}{Astrophys. J.} \textbf{\bibinfo{volume}{725}},
  \bibinfo{pages}{1563} (\bibinfo{year}{2010}).

\bibitem[{\citenamefont{Hanke et~al.}(2013)\citenamefont{Hanke, Mueller,
  Wongwathanarat, Marek, and Janka}}]{Hanke:2013jat}
\bibinfo{author}{\bibfnamefont{F.}~\bibnamefont{Hanke}},
  \bibinfo{author}{\bibfnamefont{B.}~\bibnamefont{Mueller}},
  \bibinfo{author}{\bibfnamefont{A.}~\bibnamefont{Wongwathanarat}},
  \bibinfo{author}{\bibfnamefont{A.}~\bibnamefont{Marek}}, \bibnamefont{and}
  \bibinfo{author}{\bibfnamefont{H.-T.} \bibnamefont{Janka}},
  \bibinfo{journal}{Astrophys. J.} \textbf{\bibinfo{volume}{770}},
  \bibinfo{pages}{66} (\bibinfo{year}{2013}).

\bibitem[{\citenamefont{O'Connor and Couch}(2018)}]{OConnor:2018tuw}
\bibinfo{author}{\bibfnamefont{E.~P.} \bibnamefont{O'Connor}} \bibnamefont{and}
  \bibinfo{author}{\bibfnamefont{S.~M.} \bibnamefont{Couch}},
  \bibinfo{journal}{Astrophys. J.} \textbf{\bibinfo{volume}{865}},
  \bibinfo{pages}{81} (\bibinfo{year}{2018}).

\bibitem[{\citenamefont{Vartanyan et~al.}(2019)\citenamefont{Vartanyan,
  Burrows, and Radice}}]{Vartanyan:2019ssu}
\bibinfo{author}{\bibfnamefont{D.}~\bibnamefont{Vartanyan}},
  \bibinfo{author}{\bibfnamefont{A.}~\bibnamefont{Burrows}}, \bibnamefont{and}
  \bibinfo{author}{\bibfnamefont{D.}~\bibnamefont{Radice}},
  \bibinfo{journal}{Mon. Not. Roy. Astron. Soc.}
  \textbf{\bibinfo{volume}{489}}, \bibinfo{pages}{2227} (\bibinfo{year}{2019}).

\bibitem[{\citenamefont{Walk et~al.}(2019)\citenamefont{Walk, Tamborra, Janka,
  and Summa}}]{Walk:2019miz}
\bibinfo{author}{\bibfnamefont{L.}~\bibnamefont{Walk}},
  \bibinfo{author}{\bibfnamefont{I.}~\bibnamefont{Tamborra}},
  \bibinfo{author}{\bibfnamefont{H.-T.} \bibnamefont{Janka}}, \bibnamefont{and}
  \bibinfo{author}{\bibfnamefont{A.}~\bibnamefont{Summa}}
  (\bibinfo{year}{2019}), \eprint{arXiv: 1910.12971}.

\bibitem[{\citenamefont{Tamborra et~al.}(2013)\citenamefont{Tamborra, Hanke,
  M{\"u}ller, Janka, and Raffelt}}]{Tamborra:2013laa}
\bibinfo{author}{\bibfnamefont{I.}~\bibnamefont{Tamborra}},
  \bibinfo{author}{\bibfnamefont{F.}~\bibnamefont{Hanke}},
  \bibinfo{author}{\bibfnamefont{B.}~\bibnamefont{M{\"u}ller}},
  \bibinfo{author}{\bibfnamefont{H.-T.} \bibnamefont{Janka}}, \bibnamefont{and}
  \bibinfo{author}{\bibfnamefont{G.}~\bibnamefont{Raffelt}},
  \bibinfo{journal}{Phys. Rev. Lett.} \textbf{\bibinfo{volume}{111}},
  \bibinfo{pages}{121104} (\bibinfo{year}{2013}).

\bibitem[{\citenamefont{Walk et~al.}(2018)\citenamefont{Walk, Tamborra, Janka,
  and Summa}}]{Walk:2018gaw}
\bibinfo{author}{\bibfnamefont{L.}~\bibnamefont{Walk}},
  \bibinfo{author}{\bibfnamefont{I.}~\bibnamefont{Tamborra}},
  \bibinfo{author}{\bibfnamefont{H.-T.} \bibnamefont{Janka}}, \bibnamefont{and}
  \bibinfo{author}{\bibfnamefont{A.}~\bibnamefont{Summa}},
  \bibinfo{journal}{Phys. Rev.} \textbf{\bibinfo{volume}{D98}},
  \bibinfo{pages}{123001} (\bibinfo{year}{2018}).

\bibitem[{\citenamefont{Lund et~al.}(2010)\citenamefont{Lund, Marek, Lunardini,
  Janka, and Raffelt}}]{Lund:2010kh}
\bibinfo{author}{\bibfnamefont{T.}~\bibnamefont{Lund}},
  \bibinfo{author}{\bibfnamefont{A.}~\bibnamefont{Marek}},
  \bibinfo{author}{\bibfnamefont{C.}~\bibnamefont{Lunardini}},
  \bibinfo{author}{\bibfnamefont{H.-T.} \bibnamefont{Janka}}, \bibnamefont{and}
  \bibinfo{author}{\bibfnamefont{G.}~\bibnamefont{Raffelt}},
  \bibinfo{journal}{Phys. Rev.} \textbf{\bibinfo{volume}{D82}},
  \bibinfo{pages}{063007} (\bibinfo{year}{2010}).

\bibitem[{\citenamefont{Beacom and Vogel}(1998)}]{Beacom:1998yb}
\bibinfo{author}{\bibfnamefont{J.~F.} \bibnamefont{Beacom}} \bibnamefont{and}
  \bibinfo{author}{\bibfnamefont{P.}~\bibnamefont{Vogel}},
  \bibinfo{journal}{Phys. Rev.} \textbf{\bibinfo{volume}{D58}},
  \bibinfo{pages}{093012} (\bibinfo{year}{1998}).

\bibitem[{\citenamefont{Rampp and Janka}(2002)}]{Rampp:2002bq}
\bibinfo{author}{\bibfnamefont{M.}~\bibnamefont{Rampp}} \bibnamefont{and}
  \bibinfo{author}{\bibfnamefont{H.~T.} \bibnamefont{Janka}},
  \bibinfo{journal}{Astron. Astrophys.} \textbf{\bibinfo{volume}{396}},
  \bibinfo{pages}{361} (\bibinfo{year}{2002}).

\bibitem[{\citenamefont{Buras et~al.}(2006)\citenamefont{Buras, Rampp, Janka,
  and Kifonidis}}]{Buras:2005rp}
\bibinfo{author}{\bibfnamefont{R.}~\bibnamefont{Buras}},
  \bibinfo{author}{\bibfnamefont{M.}~\bibnamefont{Rampp}},
  \bibinfo{author}{\bibfnamefont{H.~T.} \bibnamefont{Janka}}, \bibnamefont{and}
  \bibinfo{author}{\bibfnamefont{K.}~\bibnamefont{Kifonidis}},
  \bibinfo{journal}{Astron. Astrophys.} \textbf{\bibinfo{volume}{447}},
  \bibinfo{pages}{1049} (\bibinfo{year}{2006}).

\bibitem[{\citenamefont{Mueller et~al.}(2013)\citenamefont{Mueller, Janka, and
  Marek}}]{Mueller:2012sv}
\bibinfo{author}{\bibfnamefont{B.}~\bibnamefont{Mueller}},
  \bibinfo{author}{\bibfnamefont{H.-T.} \bibnamefont{Janka}}, \bibnamefont{and}
  \bibinfo{author}{\bibfnamefont{A.}~\bibnamefont{Marek}},
  \bibinfo{journal}{Astrophys. J.} \textbf{\bibinfo{volume}{766}},
  \bibinfo{pages}{43} (\bibinfo{year}{2013}).

\bibitem[{\citenamefont{Lund et~al.}(2012)\citenamefont{Lund, Wongwathanarat,
  Janka, Muller, and Raffelt}}]{Lund:2012vm}
\bibinfo{author}{\bibfnamefont{T.}~\bibnamefont{Lund}},
  \bibinfo{author}{\bibfnamefont{A.}~\bibnamefont{Wongwathanarat}},
  \bibinfo{author}{\bibfnamefont{H.-T.} \bibnamefont{Janka}},
  \bibinfo{author}{\bibfnamefont{E.}~\bibnamefont{Muller}}, \bibnamefont{and}
  \bibinfo{author}{\bibfnamefont{G.}~\bibnamefont{Raffelt}},
  \bibinfo{journal}{Phys. Rev.} \textbf{\bibinfo{volume}{D86}},
  \bibinfo{pages}{105031} (\bibinfo{year}{2012}).

\bibitem[{\citenamefont{Migenda}(2016)}]{Migenda:2016xnc}
\bibinfo{author}{\bibfnamefont{J.}~\bibnamefont{Migenda}}, Ph.D. thesis,
  \bibinfo{school}{Munich, Max Planck Inst.} (\bibinfo{year}{2016}),
  \eprint{arXiv: 1609.04286}.

\bibitem[{\citenamefont{Tamborra et~al.}(2014)\citenamefont{Tamborra, Hanke,
  Janka, Müller, Raffelt, and Marek}}]{Tamborra:2014aua}
\bibinfo{author}{\bibfnamefont{I.}~\bibnamefont{Tamborra}},
  \bibinfo{author}{\bibfnamefont{F.}~\bibnamefont{Hanke}},
  \bibinfo{author}{\bibfnamefont{H.-T.} \bibnamefont{Janka}},
  \bibinfo{author}{\bibfnamefont{B.}~\bibnamefont{Müller}},
  \bibinfo{author}{\bibfnamefont{G.~G.} \bibnamefont{Raffelt}},
  \bibnamefont{and} \bibinfo{author}{\bibfnamefont{A.}~\bibnamefont{Marek}},
  \bibinfo{journal}{Astrophys. J.} \textbf{\bibinfo{volume}{792}},
  \bibinfo{pages}{96} (\bibinfo{year}{2014}).

\bibitem[{\citenamefont{Abe et~al.}(2018)}]{Abe:2018uyc}
\bibinfo{author}{\bibfnamefont{K.}~\bibnamefont{Abe}} \bibnamefont{et~al.}
  (\bibinfo{collaboration}{Hyper-Kamiokande}) (\bibinfo{year}{2018}),
  \eprint{arXiv: 1805.04163}.

\bibitem[{\citenamefont{Scholberg}(2012)}]{Scholberg:2012id}
\bibinfo{author}{\bibfnamefont{K.}~\bibnamefont{Scholberg}},
  \bibinfo{journal}{Ann. Rev. Nucl. Part. Sci.} \textbf{\bibinfo{volume}{62}},
  \bibinfo{pages}{81} (\bibinfo{year}{2012}).

\bibitem[{\citenamefont{K{\"o}pke}(2018)}]{Kopke:2017req}
\bibinfo{author}{\bibfnamefont{L.}~\bibnamefont{K{\"o}pke}}
  (\bibinfo{collaboration}{IceCube}), \bibinfo{journal}{J. Phys. Conf. Ser.}
  \textbf{\bibinfo{volume}{1029}}, \bibinfo{pages}{012001}
  (\bibinfo{year}{2018}).

\bibitem[{\citenamefont{Kowarik et~al.}(2009)\citenamefont{Kowarik, Griesel,
  and Piegsa}}]{Kowarik:2009qr}
\bibinfo{author}{\bibfnamefont{T.}~\bibnamefont{Kowarik}},
  \bibinfo{author}{\bibfnamefont{T.}~\bibnamefont{Griesel}}, \bibnamefont{and}
  \bibinfo{author}{\bibfnamefont{A.}~\bibnamefont{Piegsa}}
  (\bibinfo{collaboration}{IceCube}) (\bibinfo{year}{2009}), \eprint{arXiv:
  0908.0441}.

\bibitem[{\citenamefont{Steiner et~al.}(2013)\citenamefont{Steiner, Hempel, and
  Fischer}}]{Steiner:2012rk}
\bibinfo{author}{\bibfnamefont{A.~W.} \bibnamefont{Steiner}},
  \bibinfo{author}{\bibfnamefont{M.}~\bibnamefont{Hempel}}, \bibnamefont{and}
  \bibinfo{author}{\bibfnamefont{T.}~\bibnamefont{Fischer}},
  \bibinfo{journal}{Astrophys. J.} \textbf{\bibinfo{volume}{774}},
  \bibinfo{pages}{17} (\bibinfo{year}{2013}).

\bibitem[{\citenamefont{Woosley and Weaver}(1995)}]{Woosley:1995ip}
\bibinfo{author}{\bibfnamefont{S.~E.} \bibnamefont{Woosley}} \bibnamefont{and}
  \bibinfo{author}{\bibfnamefont{T.~A.} \bibnamefont{Weaver}},
  \bibinfo{journal}{Astrophys. J. Suppl.} \textbf{\bibinfo{volume}{101}},
  \bibinfo{pages}{181} (\bibinfo{year}{1995}).

\bibitem[{\citenamefont{Yakunin et~al.}(2017)\citenamefont{Yakunin, Mezzacappa,
  Marronetti, Lentz, Bruenn, Hix, Bronson~Messer, Endeve, Blondin, and
  Harris}}]{Yakunin:2017tus}
\bibinfo{author}{\bibfnamefont{K.~N.} \bibnamefont{Yakunin}},
  \bibinfo{author}{\bibfnamefont{A.}~\bibnamefont{Mezzacappa}},
  \bibinfo{author}{\bibfnamefont{P.}~\bibnamefont{Marronetti}},
  \bibinfo{author}{\bibfnamefont{E.~J.} \bibnamefont{Lentz}},
  \bibinfo{author}{\bibfnamefont{S.~W.} \bibnamefont{Bruenn}},
  \bibinfo{author}{\bibfnamefont{W.~R.} \bibnamefont{Hix}},
  \bibinfo{author}{\bibfnamefont{O.~E.} \bibnamefont{Bronson~Messer}},
  \bibinfo{author}{\bibfnamefont{E.}~\bibnamefont{Endeve}},
  \bibinfo{author}{\bibfnamefont{J.~M.} \bibnamefont{Blondin}},
  \bibnamefont{and} \bibinfo{author}{\bibfnamefont{J.~A.} \bibnamefont{Harris}}
  (\bibinfo{year}{2017}).

\bibitem[{\citenamefont{Kuroda et~al.}(2016)\citenamefont{Kuroda, Kotake, and
  Takiwaki}}]{Kuroda:2016bjd}
\bibinfo{author}{\bibfnamefont{T.}~\bibnamefont{Kuroda}},
  \bibinfo{author}{\bibfnamefont{K.}~\bibnamefont{Kotake}}, \bibnamefont{and}
  \bibinfo{author}{\bibfnamefont{T.}~\bibnamefont{Takiwaki}},
  \bibinfo{journal}{Astrophys. J.} \textbf{\bibinfo{volume}{829}},
  \bibinfo{pages}{L14} (\bibinfo{year}{2016}).

\bibitem[{\citenamefont{Shibata and Nakamura}(1995)}]{Shibata:1995we}
\bibinfo{author}{\bibfnamefont{M.}~\bibnamefont{Shibata}} \bibnamefont{and}
  \bibinfo{author}{\bibfnamefont{T.}~\bibnamefont{Nakamura}},
  \bibinfo{journal}{Phys. Rev.} \textbf{\bibinfo{volume}{D52}},
  \bibinfo{pages}{5428} (\bibinfo{year}{1995}).

\bibitem[{\citenamefont{Baumgarte and Shapiro}(1998)}]{Baumgarte:1998te}
\bibinfo{author}{\bibfnamefont{T.~W.} \bibnamefont{Baumgarte}}
  \bibnamefont{and} \bibinfo{author}{\bibfnamefont{S.~L.}
  \bibnamefont{Shapiro}}, \bibinfo{journal}{Phys. Rev.}
  \textbf{\bibinfo{volume}{D59}}, \bibinfo{pages}{024007}
  (\bibinfo{year}{1998}).

\bibitem[{\citenamefont{Kuroda et~al.}(2012)\citenamefont{Kuroda, Kotake, and
  Takiwaki}}]{Kuroda:2012nc}
\bibinfo{author}{\bibfnamefont{T.}~\bibnamefont{Kuroda}},
  \bibinfo{author}{\bibfnamefont{K.}~\bibnamefont{Kotake}}, \bibnamefont{and}
  \bibinfo{author}{\bibfnamefont{T.}~\bibnamefont{Takiwaki}},
  \bibinfo{journal}{Astrophys. J.} \textbf{\bibinfo{volume}{755}},
  \bibinfo{pages}{11} (\bibinfo{year}{2012}).

\bibitem[{\citenamefont{Mikheyev and Smirnov}(1985)}]{Mikheev:1986gs}
\bibinfo{author}{\bibfnamefont{S.~P.} \bibnamefont{Mikheyev}} \bibnamefont{and}
  \bibinfo{author}{\bibfnamefont{A.~{\relax Yu}.} \bibnamefont{Smirnov}},
  \bibinfo{journal}{Sov. J. Nucl. Phys.} \textbf{\bibinfo{volume}{42}},
  \bibinfo{pages}{913} (\bibinfo{year}{1985}), \bibinfo{note}{[,305(1986)]}.

\bibitem[{\citenamefont{Kay}(2009)}]{2009fundamentals}
\bibinfo{author}{\bibfnamefont{S.~M.} \bibnamefont{Kay}},
  \emph{\bibinfo{title}{Fundamentals Of Statistical Processing, Volume 2:
  Detection Theory}}, Prentice-Hall signal processing series
  (\bibinfo{publisher}{Pearson Education}, \bibinfo{year}{2009}), ISBN
  \bibinfo{isbn}{9788131729007},
  \urlprefix\url{https://books.google.com/books?id=wwmnY9xyt9MC}.

\bibitem[{\citenamefont{Radice et~al.}(2019)\citenamefont{Radice, Morozova,
  Burrows, Vartanyan, and Nagakura}}]{Radice:2018usf}
\bibinfo{author}{\bibfnamefont{D.}~\bibnamefont{Radice}},
  \bibinfo{author}{\bibfnamefont{V.}~\bibnamefont{Morozova}},
  \bibinfo{author}{\bibfnamefont{A.}~\bibnamefont{Burrows}},
  \bibinfo{author}{\bibfnamefont{D.}~\bibnamefont{Vartanyan}},
  \bibnamefont{and} \bibinfo{author}{\bibfnamefont{H.}~\bibnamefont{Nagakura}},
  \bibinfo{journal}{Astrophys. J.} \textbf{\bibinfo{volume}{876}},
  \bibinfo{pages}{L9} (\bibinfo{year}{2019}).

\bibitem[{\citenamefont{Powell and Müller}(2019)}]{Powell:2018isq}
\bibinfo{author}{\bibfnamefont{J.}~\bibnamefont{Powell}} \bibnamefont{and}
  \bibinfo{author}{\bibfnamefont{B.}~\bibnamefont{Müller}},
  \bibinfo{journal}{Mon. Not. Roy. Astron. Soc.}
  \textbf{\bibinfo{volume}{487}}, \bibinfo{pages}{1178} (\bibinfo{year}{2019}).

\bibitem[{\citenamefont{M{\"u}ller et~al.}(2012)\citenamefont{M{\"u}ller,
  Janka, and Wongwathanarat}}]{Muller:2011yi}
\bibinfo{author}{\bibfnamefont{E.}~\bibnamefont{M{\"u}ller}},
  \bibinfo{author}{\bibfnamefont{H.~T.} \bibnamefont{Janka}}, \bibnamefont{and}
  \bibinfo{author}{\bibfnamefont{A.}~\bibnamefont{Wongwathanarat}},
  \bibinfo{journal}{Astron. Astrophys.} \textbf{\bibinfo{volume}{537}},
  \bibinfo{pages}{A63} (\bibinfo{year}{2012}).

\bibitem[{\citenamefont{Mueller et~al.}(2004)\citenamefont{Mueller, Rampp,
  Buras, Janka, and Shoemaker}}]{Mueller:2003fs}
\bibinfo{author}{\bibfnamefont{E.}~\bibnamefont{Mueller}},
  \bibinfo{author}{\bibfnamefont{M.}~\bibnamefont{Rampp}},
  \bibinfo{author}{\bibfnamefont{R.}~\bibnamefont{Buras}},
  \bibinfo{author}{\bibfnamefont{H.~T.} \bibnamefont{Janka}}, \bibnamefont{and}
  \bibinfo{author}{\bibfnamefont{D.~H.} \bibnamefont{Shoemaker}},
  \bibinfo{journal}{Astrophys. J.} \textbf{\bibinfo{volume}{603}},
  \bibinfo{pages}{221} (\bibinfo{year}{2004}).

\bibitem[{\citenamefont{Murphy et~al.}(2009)\citenamefont{Murphy, Ott, and
  Burrows}}]{Murphy:2009dx}
\bibinfo{author}{\bibfnamefont{J.~W.} \bibnamefont{Murphy}},
  \bibinfo{author}{\bibfnamefont{C.~D.} \bibnamefont{Ott}}, \bibnamefont{and}
  \bibinfo{author}{\bibfnamefont{A.}~\bibnamefont{Burrows}},
  \bibinfo{journal}{Astrophys. J.} \textbf{\bibinfo{volume}{707}},
  \bibinfo{pages}{1173} (\bibinfo{year}{2009}).

\bibitem[{\citenamefont{Yakunin et~al.}(2010)}]{Yakunin:2010fn}
\bibinfo{author}{\bibfnamefont{K.~N.} \bibnamefont{Yakunin}}
  \bibnamefont{et~al.}, \bibinfo{journal}{Class. Quant. Grav.}
  \textbf{\bibinfo{volume}{27}}, \bibinfo{pages}{194005}
  (\bibinfo{year}{2010}).

\bibitem[{\citenamefont{Blackburn et~al.}(2005)}]{Blackburn:2005qv}
\bibinfo{author}{\bibfnamefont{L.}~\bibnamefont{Blackburn}}
  \bibnamefont{et~al.}, \bibinfo{journal}{Class. Quant. Grav.}
  \textbf{\bibinfo{volume}{22}}, \bibinfo{pages}{S1293} (\bibinfo{year}{2005}),
  \eprint{gr-qc/0504060}.

\bibitem[{\citenamefont{Aartsen et~al.}(2018)}]{IceCube:2018dnn}
\bibinfo{author}{\bibfnamefont{M.~G.} \bibnamefont{Aartsen}}
  \bibnamefont{et~al.} (\bibinfo{collaboration}{IceCube, Fermi-LAT, MAGIC,
  AGILE, ASAS-SN, HAWC, H.E.S.S., INTEGRAL, Kanata, Kiso, Kapteyn, Liverpool
  Telescope, Subaru, Swift NuSTAR, VERITAS, VLA/17B-403}),
  \bibinfo{journal}{Science} \textbf{\bibinfo{volume}{361}},
  \bibinfo{pages}{eaat1378} (\bibinfo{year}{2018}).

\bibitem[{\citenamefont{Branchesi}(2016)}]{Branchesi:2016vef}
\bibinfo{author}{\bibfnamefont{M.}~\bibnamefont{Branchesi}},
  \bibinfo{journal}{J. Phys. Conf. Ser.} \textbf{\bibinfo{volume}{718}},
  \bibinfo{pages}{022004} (\bibinfo{year}{2016}).

\bibitem[{\citenamefont{Kalogera et~al.}(2019)}]{Kalogera:2019bdd}
\bibinfo{author}{\bibfnamefont{V.}~\bibnamefont{Kalogera}} \bibnamefont{et~al.}
  (\bibinfo{year}{2019}), \eprint{arXiv: 1903.09224}.

\end{thebibliography}


\end{document}